# Emergent Atomic Scale Polarisation Vortices in $BaTiS_3$

Boyang Zhao[1,2]†, Gwan Yeong Jung[3]†, Shantanu Singh[1], Robert B. Smith[4,5], Huandong Chen[1], Guodong Ren[6], Chuangtang Wang[7], Sara Termos[4,5], Sean T. Holmes[4,5], Frederic Mentink-Vigier[4,5], Weizhe Zhang[7], Zhengyu Du[1], Claire Wu[1], M. J. Swamynadhan[3], Qinai Zhao[1], Kevin Ye[1], Donald A. Walko[8], Nicholas S. Settineri[9], Simon J. Teat[9], Liuyan Zhao[7], Robert W. Schurko[4,5], Haidan Wen[2,8], Rohan Mishra[3,6,*], and Jayakanth Ravichandran[1,10,11,*]

[1]*Mork Family Department of Chemical Engineering and Materials Science, University of Southern California, Los Angeles, CA 90089, USA*

[2]*Materials Science Division, Argonne National Laboratory, Lemont, IL 60439, USA*

[3]*Department of Mechanical Engineering & Materials Science, Washington University in St. Louis, St. Louis, MO 63130, USA*

[4]*Department of Chemistry and Biochemistry, Florida State University, Tallahassee, FL 32306, United States*

[5]*National High Magnetic Field Laboratory, Tallahassee, Florida 32310, United States*

[6]*Institute of Materials Science & Engineering, Washington University in St. Louis, St. Louis, MO 63130, USA*

[7]*Department of Physics, University of Michigan, Ann Arbor, MI 48019, USA*

[8]*Advanced Photon Source Argonne National Laboratory, Lemont, IL 60439, USA*

[9]*Advanced Light Source, Lawrence Berkeley National Laboratory; Berkeley, CA 94720, USA*

[10]*Ming Hsieh Department of Electrical Engineering, University of Southern California; Los Angeles, CA 90089, USA*

[11]*Core Center of Excellence in Nano Imaging, University of Southern California; Los Angeles, CA 90089, USA*

† These authors contributed equally: Boyang Zhao, Gwan Yeong Jung

* Corresponding author. Email: j.ravichandran@usc.edu, rmishra@wustl.edu

**Topological defects, such as vortices and skyrmions in magnetic and dipolar systems, can give rise to properties that are not observed in typical magnets and dielectrics. Here, we report the discovery of long-range ordered periodic dipole arrays of atomic-scale vortices and antivortices in the unconventional charge-density-wave (CDW) phase of $BaTiS_3$, a quasi-1D chalcogenide. Synchrotron X-ray diffraction (XRD) reveals the presence of a multi-*q* ordering in $BaTiS_3$ that confines vortex–vortex–antivortex polarisation triplets to the *a-b* plane with alternating handedness along the *c*-axis. The multi-*q* displacive distortions are characterised by three distinctive off-centre $TiS_6$ configurations, whose ratios are independently confirmed by $^{47/49}Ti$ solid-state nuclear magnetic resonance (SSNMR). Using first-principles calculations and phenomenological modelling, we show that the dipolar vortex unit cell in $BaTiS_3$ arises from the coupling between multiple lattice instabilities arising from flat, *soft* phonon bands. This mechanism contrasts with classical dipolar textures in ferroelectric heterostructures that emerge from the competition between electrostatic and strain energies. The observation of dipolar vortices in $BaTiS_3$ brings the ultimate scaling limit for real-space dipolar topological structures down to about a nanometre and unveils the intimate connection between crystal symmetry and real-space topology. Our work sets up zero-filling semiconducting materials with competing structural instabilities as a playground for realising and understanding quantum polarisation topologies.**

## Introduction

Topological defects are a vibrant area of research, extending from cosmology[1,2] to condensed matter physics[3,4], as they uncover singularities within the ordered states of systems. The scaling limits of such defects have typically remained in the mesoscale (*e.g.* liquid crystals[5]), but advances in atomic and nano-scale synthetic control and characterisation probes[6,7] for spins[8] and electric dipoles[9] have unveiled nanoscale topological defects or structures such as vortices and skyrmions in magnetic systems, and more recently, in low-dimensional ferroelectrics[9,10]. Such nanoscale non-collinear textures in both magnetic and dipolar systems hold significance for information storage and processing[11]. Hence, understanding the scaling limit of these structures is of fundamental importance. In magnetic systems, the size of spin textures commonly ranges from tens to hundreds of nanometres, depending on the dominant exchange interaction[12,13]. Recently, the studies of temperature-modulated spin orders in Kagome lattices revealed non-trivial spin-textures[14], such as Kagome spin-ice[15] in frustrated 2D antiferromagnets, lowering the size of long-range magnetic textures to the "atomic-scale", or the scale of a superstructure unit cell.

In contrast, the size of non-collinear topological textures in dipolar systems, so far, is dictated by the geometrical constraints that can be imposed in low-dimensional ferroelectrics by changing the film thickness, engineering heterostructures[10], or through the growth of nanowires[16]. Here, the competition between long-range interactions, represented by electrostatic and strain energy, and the short-range polarisation gradient energy, dictates the stability and scale of the dipolar textures. Thus, stabilisation of the short-range dipolar textures in low-dimensional ferroelectrics requires careful tuning of the long-range boundary conditions imposed by the geometry and the strain state[9,17]. Furthermore, strong depolarisation effects at the ultralow thicknesses[18] preclude the formation of such topological textures, thereby placing a lower bound on their size to several unit cells[19,20], *i.e.*, > 4 nm in the case of the widely studied $SrTiO_3/PbTiO_3$ system. This naturally raises the question of whether a fundamentally different mechanism can overcome this limitation and give rise to atomic-scale topological textures.

The similarities between electricity and magnetism have prompted the hypothesis of asymmetric-exchange-type interactions in dipolar systems[21,22]. Coupled lattice instabilities have been theoretically proposed as a pathway to non-collinear dipolar textures, potentially at smaller length scales than the classical dipolar textures[23–25]. Analogous to ferroelectric materials, charge density wave (CDW) or charge ordered materials, such as 1*T*-$TiSe_2$[26] and $KV_3Sb_5$[27], also harbour non-collinear lattice instabilities. Their coupled reciprocal space wave vectors stabilise a toroidal, or sometimes chiral, charge-density-wave-like order, expanding the unit cell with non-collinear displacive distortions, and create a pathway to realise topological dipolar order at a length scale much smaller than the dipolar textures in classic ferroelectric heterostructures.

In this article, we report the experimental observation of atomic-scale non-collinear dipolar textures due to coupled lattice instabilities in the trigonal phase of a quasi-1D semiconducting material, $BaTiS_3$, using a combination of single crystal X-ray diffraction (SC-XRD), rotation anisotropy second harmonic generation (RA-SHG), and solid-state nuclear magnetic resonance (SSNMR). In addition to the antiparallel dipole orders along the *c*-axis, the non-collinear *a*-*b* plane dipolar displacements of $TiS_6$ octahedra in $BaTiS_3$ form a superstructure unit cell with topologically non-trivial *a*-*b* plane polarisation vortices, resembling a 2D Kagome lattice-like structure of clockwise, anticlockwise, head-to-head and tail-to-tail vortex triplets, which reverse the vortex handedness every half-unit cell along the *c*-axis. Using first-principles density-functional theory (DFT) calculations, we show that the observed lattice instabilities originate from coupled *soft* modes in the phonon band structure of the high-symmetry phase of $BaTiS_3$. Using a Landau model built with inputs from DFT total energies, we verified that the coupling of several soft phonon modes gives rise to the intricate vortex unit cell of $BaTiS_3$.

## Phase Change and Lattice Instabilities of $BaTiS_3$

$BaTiS_3$ adopts a $BaNiO_3$-type hexagonal lattice structure with face-shared $TiS_6$ octahedra forming quasi-one-dimensional (quasi-1D) chains along the *c*-axis (**Figure 1a**). Originally refined into a centrosymmetric $P6_3/mmc$ crystal structure *via* powder X-ray diffraction (PXRD)[28], recent diffraction studies on high-quality single crystals of $BaTiS_3$ revealed the discovery of the $\Gamma_2^-$-type ($P6_3mc$ unit cell[29,30]) $TiS_6$ dipolar displacements, which couple with the $K_3$-type ($P6_3cm$ unit cell[31,32]) antiparallel displacive distortions from an *in-depth* room-temperature structure solution [**Extended Data Figure 1**]. The global displacements of quasi-1D $TiS_6$ chains along the *c*-axis are combined with antiparallel off-centre Ti displacements (~ 0.2 Å) of the $TiS_6$ octahedra, which are comparable in magnitude to the dipolar displacements observed in prominent ferroelectric perovskites such as $BaTiO_3$[33] and $PbTiO_3$[34]. These dipolar displacements in $BaTiS_3$ play a critical role in interpreting its unique optical[29,32], electrical[31,35], and thermal[30,36] properties. Diffraction studies with X-rays, particularly those utilising the highly brilliant and monochromatic X-rays produced by synchrotron radiation, reveal the long-range displacive distortions as a series of non-integer *hkl* Bragg peaks in reciprocal space, providing sensitivity to subtle yet ordered displacements with accurate projections in the extended reciprocal space. Furthermore, the exceptional synchrotron beam stability and the feasibility of using a cryo-stream for transmission geometry XRD on pristine $BaTiS_3$ crystal enable the reconstruction of the reciprocal space at cryogenic temperatures with unparalleled precision. Therefore, we focus on studying the ordering of dipolar displacements in $BaTiS_3$ using extended reciprocal space X-ray diffraction.

$BaTiS_3$ undergoes two phase transitions at cryogenic conditions: (I) a first-order structural transition between ~ 150 - 190 K, and (II) an unconventional zero-filling CDW transition with coincident electronic and structural transition[31,35] between ~ 240 - 260 K (**Figure 1b**), but little attention has been paid to the emergent dipole order in these phases. In **Figure 1c**, we re-examine the displacive distortions of the CDW phase, with the integrated *hk*1 and *hk*2 precession maps (cross-section of the reciprocal space map through *hk*1 and *hk*2 reflections) of the single crystal X-ray diffraction (SC-XRD) measurements[37,38] on $BaTiS_3$ at 220 K (more details in **Methods III** and **Supplementary Materials, Section 2**). The primary reflections, characterised by integers *h*, *k*, *l* at the centre of the Brillouin zone, originate from the high-symmetry parent structure of $BaTiS_3$ (space group: $P6_3/mmc$), across all temperatures. In contrast, the satellite reflections reveal three frozen wave vectors within the Brillouin zone: $q_1$ = (1/3, 1/3, 0) at K points, $q_2$ = (1/2, 0, 0) at M points, and $q_3$ = (1/6, 1/6, 0) at Λ points, as shown in **Figure 1d**. Unlike the room temperature phase of $BaTiS_3$ (space group: $P6_3cm$), which has only $q_1$ breaking the symmetry and extending the Brillion zone, or the low-temperature phase of $BaTiS_3$ below 150 K (space group: $P2_1$), which adopts a $q_2$ dominated superstructure, the CDW phase of $BaTiS_3$ (space group: $P3c1$) has coexisting $q_1$, $q_2$, and $q_3$ (**Figure 1d**) that show a collective, and smooth temperature evolution during both the nucleation and the melting of this phase (**Figure 1b**). More interestingly, when examining the peak intensities corresponding to $q_1$, $q_2$, and $q_3$, we find that they not only adhere to triple-*q* rotational symmetry (3*m* Laue symmetry) and the *c*-glide symmetry (extinction of the *hhl* when *l* is odd), but also exhibit symmetry unrelated intensity profiles such as the $q_3$ = (1/6, 1/6, 1) and (2/6, $\overline{1/6}$, 1) orientations being ~1/10 intensity of other $q_3$ satellite reflections of the 101 reflection; more such features are described in **Extended Data Figure 2a**. This indicates the presence of a hidden displacement texture that extends beyond the global symmetry operations. Extracting this texture necessitates meticulous *phase retrieval*, as described in the **Supplementary Materials, Section 3**.

We thus employed the direct method of crystallography[39] for the phase retrieval of the structure factors. The modulation wave vectors $q_1$, $q_2$, and $q_3$ are all commensurate with the $P6_3/mmc$ $BaTiS_3$ parent structure and produce sharp satellite reflections [**Extended Data Figure 1c**]: the correlation length extracted from the peak width ($\xi = 1/\Delta q \sim 161$ nm) greatly exceeds the real-space modulation period ($1/q \sim 2.3$ nm). Therefore, we first accepted all combinations of $q_1$, $q_2$, and $q_3$ in a coupled $2\sqrt{3} \times 2\sqrt{3} \times 1$ superstructure unit cell and then used symmetry to narrow down the initial solutions. Subsequently, the structure factors of suitable solutions were refined to align with the observed diffraction patterns, $I_{hkl} \propto |F_{hkl}|^2$, up to the diffraction limit for the best results (more in **Supplementary Materials, Section 3**).

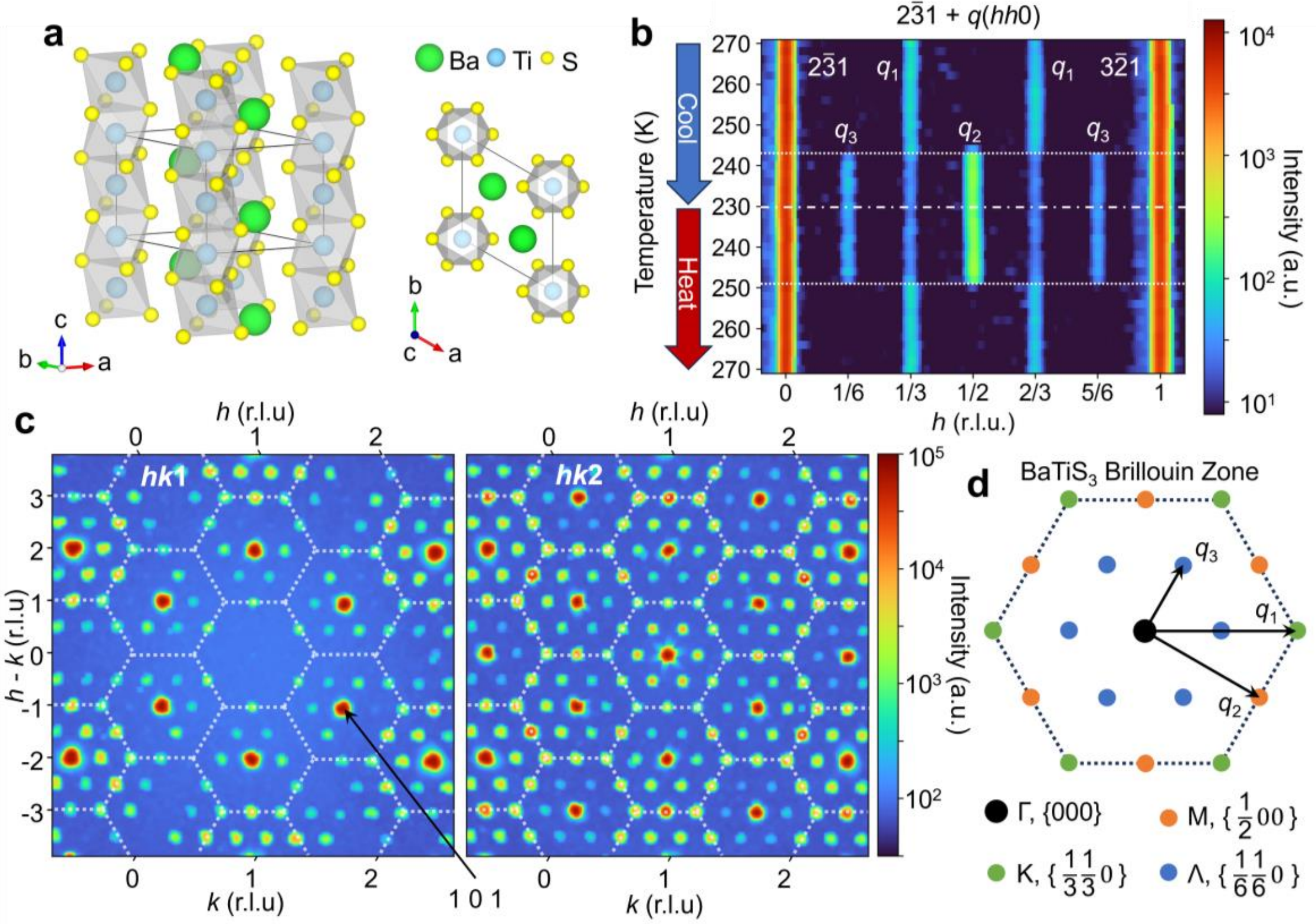

**Figure 1. Collective modulation wave vectors in BaTiS$_3$.** (**a**) Crystal structure of centrosymmetric BaTiS$_3$. (**b**) Collective structure modulations take place in the CDW phase of BaTiS$_3$. Satellite reflections at $q_2$ {1/2, 0, 0} and $q_3$ {1/6, 1/6 , 0} emerge during cooling with a concomitant reduction of the $q_1$ {1/3, 1/3, 0}, which is reversible during heating. Dotted lines mark the transition temperatures, and the dash-dot line marks the inflexion between sequential cooling and heating. (**c**) Precession maps of 220 K BaTiS$_3$ SC-XRD showing the $hk1$ and $hk2$ cross-sections of the reciprocal space $I_{hkl}$. The white dotted lines mark the Brillouin zones of the BaTiS$_3$ basic structure. (**d**) Satellite peaks are observed at K ($q_1$), M ($q_2$), and Λ ($q_3$) points within the Brillouin zone.

Beyond the reciprocal space symmetries deciphered from SC-XRD, we performed temperature-dependent rotation anisotropy second harmonic generation (RA-SHG) measurements, as a sensitive probe of symmetry breaking[40–44], on $BaTiS_3$ crystals with surface parallel to the *a*-*c* plane (of the $P6_3/mmc$ unit cell, as in the **Extended Data Figure 1**). Our chosen temperature range, from 110 K to 290 K, encompasses two phase transitions[31] across $T_{c1}$ ~ 150 - 190 K and $T_{c2}$ ~ 240 - 250 K. **Extended Data Figure 3a** shows RA-SHG polar plots recorded in the linear parallel polarisation channel at three representative temperatures, 130 K (below $T_{c1}$), 210 K (between $T_{c1}$ and $T_{c2}$), and 290 K (above $T_{c2}$). All three polar plots display a dominant two-fold symmetric pattern, with a pair of prominent lobes aligned along the *a*-axis. As the temperature increases, the RA-SHG evolves with three key changes: an overall decrease in intensity, disappearance of a nonzero background, and suppression of two smaller lobes down to zero intensity at 90°/270°. We note that the zero SHG intensity at 90°/270° at 290 K is guaranteed by the mirror plane of $P6_3cm$, and additionally that its growth into finite intensity at 210 K and even a pair of small lobes at 130 K signifies the loss of this mirror symmetry in $P3c1$ and $P2_1$, respectively.

Considering the smaller correlation length ($\xi_x$ ~ 161 nm, $\xi_z$ ~ 368 nm at 200 K in **Extended Data Figure 1e**) than the optical beam size (~2.5 μm) introduces the domain average effect, the RA-SHG patterns below $T_{c1}$ and $T_{c2}$ are seemingly mirror symmetric despite the mirror symmetry breaking in $P3c1$ and $P2_1$. **Extended Data Figure 3b** shows the temperature dependence of the SHG tensor elements arising from the broken mirror symmetry. We can clearly observe the weak first-order phase transition at ~ 227 K (lower temperature than ~ 240 – 250 K[31,35] likely due to excessive laser heating), below which the mirror breaks in the CDW $BaTiS_3$. The SC-XRD crystal structure refinement (details in Supplementary Materials Section 3) confirms these observations by showing a considerably better fit (**Extended Data Figure 4a**) when we introduce twinned domains of broken inversion/mirror symmetries, from $P6_3/mcm$ (representing $P6_3/mmc$[28] structure in the current unit cell) to $P6_3cm$ to $P3c1$.

Recent studies in magnetic systems show that distinguishing true triple-$q$ order (with 3-fold rotational symmetry) from triangularly twinned single-$q$ order (which lacks such high-symmetry rotation) has been key to identifying atomic-scale spin-ice textures[45]. In CDW-$BaTiS_3$, triple-$q$ order resides at each K, M, and Λ point, so a more meticulous multi-$q$ symmetry examination is practised for all less symmetric space groups: glide/mirror twinned $P3$, $P6_3$, 3-fold rotation twinned $Cmc2_1$, $Cc$, in comparison with their more symmetric parent space group, as shown in **Extended Data Figure 4a**. The $P3c1$ space group yields the overall lowest refinement residuals ($R_1 = 0.0202$ and $wR_2 = 0.0429$; see Supplementary Materials Section 3) as the best structural solution. It is a $2\sqrt{3} \times 2\sqrt{3} \times 1$ superstructure [**Extended Data Table 2-3**] of the $P6_3/mmc$ basic structure, or a $2 \times 2 \times 1$ superstructure of the room temperature structure. **Extended Data Figure 2b-c** shows the calculated structure factor square modulus ($|F_{hkl}|^2$) and phase ($\phi$ in $F_{hkl} = |F_{hkl}|e^{i\phi}$) of the $P3c1$ unit cell in comparison with the observed $I_{hkl} \propto |F_{hkl}|^2$ in **Extended Data Figure 2a.**

## Intricate Octahedral Dipolar Displacements

The crystal structure of CDW-$BaTiS_3$ is visualised in the [001] projection in **Figure 2a**. The symmetry elements of a CDW-$BaTiS_3$ are summarised in the **Extended Data Figures 4a-b**. Aside from the antiparallel $TiS_6$ chain displacements (concerning Ba sublattice) and $TiS_6$ octahedra dipolar displacements along the $c$-axis [**Extended Data Figure 1a**], certain $TiS_6$ octahedra also undergo intricate $a$-$b$ plane dipolar displacements, classifying dipolar $TiS_6$ octahedra into three second-order Jahn-Teller[46] (SOJT) distortions [**Extended Data Figure 5**]:

1. The symmetric $TiS_6$ octahedra, population 6 of 24, highlighted by dash-dot circles, are characterised by zero $a$-$b$ plane dipolar displacements but still susceptible to the universal $c$-axis dipolar displacements. They thus come with the same $a$-$b$ plane Ti-S distance, but different Ti-S bond lengths along the $c$-axis, an axial SOJT distortion[46].

2. The asymmetric $TiS_6$ octahedra with ~0.20 Å Ti *a-b* plane displacements toward one of the longer Ti-S, population 6 of 24, highlighted by solid circles. This distorted octahedral coordination involves two shorter and four longer Ti-S bonds, which is representative of an off-axial SOJT1 distortion[46].

3. The asymmetric $TiS_6$ octahedra featuring ~0.16 Å Ti *a-b* plane displacements toward one of the shorter Ti-S, population 12 of 24, highlighted by dotted circles, thereby forming one Ti-S bond significantly shorter than the other five Ti-S bonds, an off-axial SOJT2 distortion[46].

The spatial distribution of the SOJT-distorted $TiS_6$ octahedra is confined/controlled by the $P3c1$ symmetry and forms an intricate displacement arrangement, as in **Figure 2c**. Axial SOJT $TiS_6$ occupy the three-fold rotational axis and are constrained from possessing any *a-b* plane asymmetry; off axial SOJT1 and SOJT2 $TiS_6$, combined with the three-fold rotation, form a Kagome lattice, with one clockwise vortex, one anticlockwise vortex, and one head-to-head-tail-to-tail antivortex across each unit cell. Further, we emphasise that such vortex triplets of the dipolar displacements reverse their handedness between the upper and lower layers of the unit cell due to the half-unit-cell glide symmetry along the *c*-axis. It is worth noting that a direct imaging of the dipole order, *e.g.*, *via* high-resolution scanning transmission electron microscopy (STEM), is hindered by the intricate, alternating atomic layers along all dimensions and their subtle temperature and strain sensitivities[32,47].

In the non-polar $P6_3/mmc$ parent structure of $BaTiS_3$, the centroids of the cation and anion coincide. During phase transitions, Ti provides the dominant cation displacements relative to the parent structure, while S contributes to the anion displacements. Therefore, the relative displacement between Ti and the new centroid of $S_6$ primarily determines the polarisation of the $TiS_6$ octahedra. A direct consequence of such polarisation is the redistribution of charge next to Ti. **Extended Data Figure 6b** shows the SC-XRD electron distribution of the symmetric and the asymmetric $TiS_6$ chains. Besides the core electrons encircling the Ti nucleus, the outer-shell electrons that bond Ti and S show a clear distortion towards the Ti off-centre orientation, revealing distinctive polarisations of $TiS_6$ octahedra.

However, both the antiparallel dipolar displacements along the *c*-axis and the handedness reversals of polarisation vortices within the *a*-*b* plane minimise the net polarizability of CDW-$BaTiS_3$, effectively screening the local polarisations from being resolved by prominent polarisation-sensitive techniques such as piezoresponse force microscopy (PFM) and scanning tunnelling microscopy (STM). To overcome this limitation, we propose to validate the structure solution by probing the $TiS_6$ local symmetries using SSNMR.

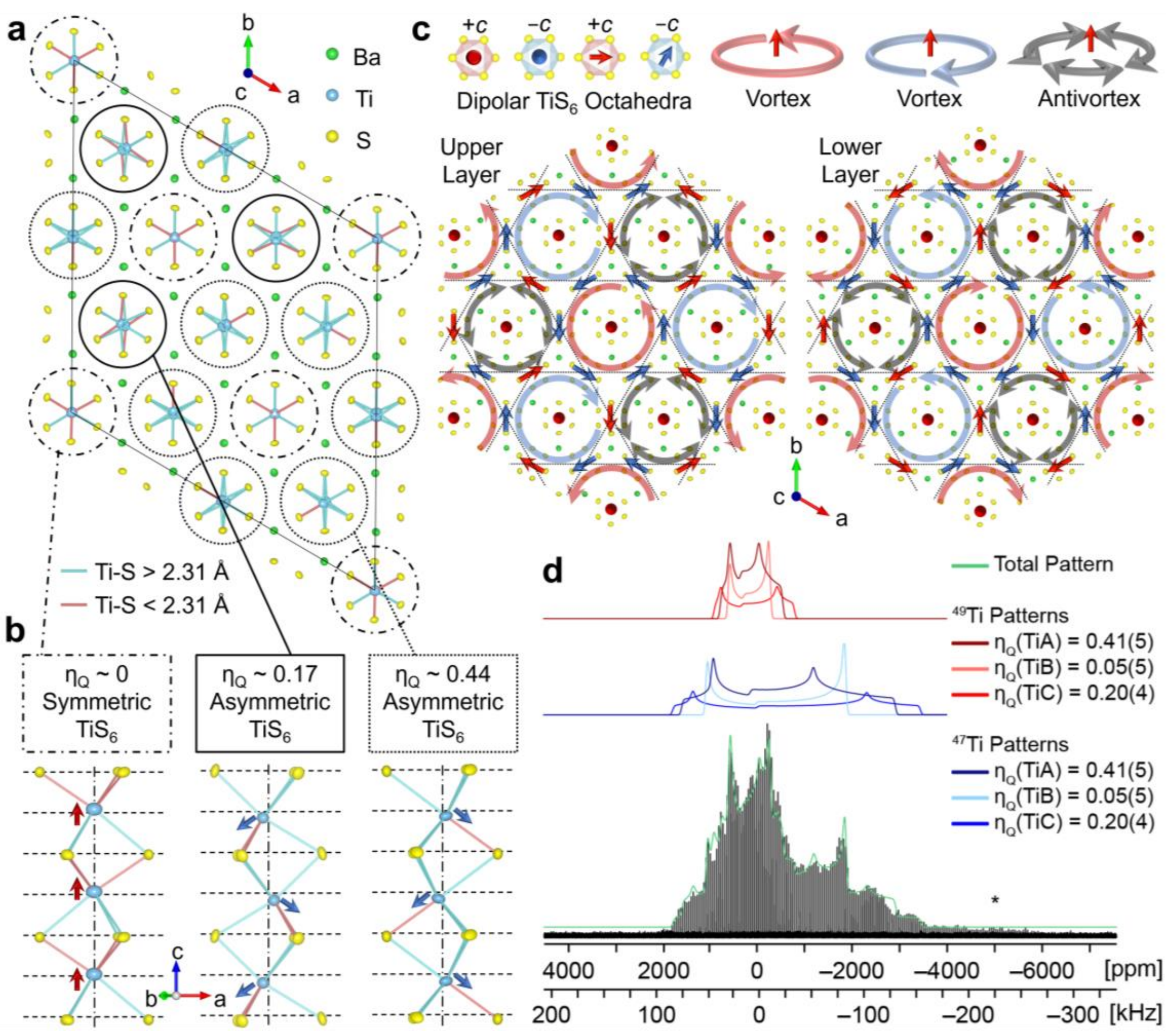

**Figure 2. Vortex unit cell of the *P*3*c*1-$BaTiS_3$.** (**a**) The *a*-*b* plane projection of the refined $BaTiS_3$ crystal structure at 220 K. Atoms are visualised as displacement ellipsoids (corresponding to the refined ADPs). Three $TiS_6$ octahedral symmetries are circled with dash-dot, solid, and dotted lines, representing symmetric,

low asymmetry, and high asymmetry $TiS_6$. Ti–S coordination is colour-coded red and blue based on bond length. (**b**) The *a-c* plane projection of some Ti-S coordination. The Ti dipolar displacements along the *a-b* plane introduce two unique zigzag dipolar modes. (**c**) Non-colinear Ti dipolar displacements (arrows represent the *a-b* plane orientation, dots represent zero *a-b* plane projection) for both the upper ($-y$, $-x$, $z + c/2$) and lower ($x$, $y$, $z$) layers of the unit cell. The asymmetric $TiS_6$ octahedra form an effective Kagome lattice, marked by dotted lines, whose spatial arrangement adopts vortex textures, colour-coded based on their handedness: blue for clockwise, red for anticlockwise, and grey for head-to-head-tail-to-tail. The *a-b* plane dipolar displacements are arranged in triplets consisting of clockwise vortex, counterclockwise vortex, and head-to-head/tail-to-tail antivortex, but of opposite handedness to the next layer, also in **Extended Data Figure 6c**. (**d**) Experimental and simulated $^{47/49}Ti$ SSNMR spectra of the CDW-$BaTiS_3$ (170 K during cooling) acquired at 18.8 T. Best fits of $^{47}Ti$ and $^{49}Ti$ central-transition ($+1/2 \leftrightarrow -1/2$) powder patterns are shaded blue and red, respectively, with the total powder pattern shown in green. Satellite-transition (*i.e.*, $m_I \leftrightarrow m_I - 1$, where $m_I = I, I - 1, \ldots, -I$) patterns are indicated by an asterisk (*). The asymmetry parameter, $\eta_Q$, describes the degree of axial symmetry of the $^{47/49}Ti$ EFG tensors (*i.e.*, $\eta_Q = 0$ indicates perfect axial symmetry, while larger values of $\eta_Q$ indicate increasing departure from axial symmetry). The best fit is obtained with three Ti sites ($\eta_Q = 0.05, 0.20, 0.41$) in a 1:1:2 population ratio, validating the Ti-S coordination order in (**a**).

The $^{47/49}Ti$ SSNMR spectra of $BaTiS_3$ provide measurements of quadrupolar coupling constants, $C_Q$, and asymmetry parameters, $\eta_Q$, which offer insights into the local symmetries of the $TiS_6$ coordination environments in each phase (See **Methods V**). For instance, the $^{47/49}Ti$ powder patterns measured at each temperature on both cooling and heating trajectories (**Extended Data Figure 7**) are consistent with the types of $TiS_6$ sites predicted by X-ray methods in each phase. In particular, the $^{47/49}Ti$ SSNMR spectra of CDW-$BaTiS_3$ (*i.e.*, with space group $P3c1$), upon cooling to 170 K, are more complex (**Figure 2d**) than those of the other two phases. Spectral simulations of the $P3c1$ phase were attempted employing one, two, and three crystallographically distinct Ti environments, each improving the overall fit by accounting for more spectral features (**Extended Data Figure 7**). The best fit involves inclusion of three Ti sites (A, B, and C) with distinct sets of quadrupolar parameters (**Figure 2d**): $C_Q^{(A)} = 12.1, 14.6$ MHz, $\eta_Q^{(A)} = 0.41$, $C_Q^{(B)} = 11.4, 13.8$ MHz, $\eta_Q^{(B)} = 0.05$, $C_Q^{(C)} = 14.8, 17.0$ MHz, $\eta_Q^{(C)} = 0.20$ (*N.B.*: the $C_Q^{(i)}$ values are reported in the format $C_Q(^{49}Ti)$, $C_Q(^{47}Ti)$, which are related by the ratio of their nuclear quadrupole moments, $Q(^{49}Ti)/Q(^{47}Ti)$). In this best fit, $^{47/49}Ti$ patterns corresponding to $\eta_Q = 0.05$, 0.20, and 0.41 are assigned integrated intensities of 1:1:2, which corresponds well with the relative populations of Wyckoff positions corresponding to symmetric, low asymmetry, and high asymmetry $TiS_6$ sites, respectively, which exist in a ratio of 6:6:12 (**Figure 2a**). These quadrupolar parameters are consistent with those obtained from plane-wave DFT calculations (**Figure 2b** and **Extended Data Table 4**), which greatly assisted in site assignment.

It is important to note that CPMG[48] (Carr-Purcell/Meiboom Gill)-type spectra are often unreliable for quantifying relative integrated intensities, as the effective transverse relaxation time constants, $T_2^{eff}$, can significantly affect the measured intensities of metal sites in different local environments. From the perspective of NMR relaxation mechanisms, the three types of $TiS_6$ sites in the $P3c1$ phase should exhibit similar $T_2^{eff}$ behaviour, since the only likely contributing relaxation mechanism is that of the quadrupolar interactions (*e.g.*, there are no heteronuclear dipolar relaxation contributions from high-γ nuclides like $^1H$ or $^{19}F$). Hence, in this case, the good fit obtained with the 1:1:2 intensities is justified. Further, this behaviour also suggests that the $^{47/49}Ti$ patterns do not arise from $TiS_6$ sites in different phases or domains, since in such instances it is highly probable that their $T_2^{eff}$ values would show greater variation. This supports the notion of the $P3c1$ displacive distortion superstructure resolved by SC-XRD. More details on the temperature dependency and the deconvolution analysis of the $^{47/49}Ti$ SSNMR spectra are given in the **Supplementary Materials, Sections 5-6**.

## Resolving the Polarisation Textures of $BaTiS_3$

To better understand the *a-b* plane dipolar displacements of *P*3*c*1 $BaTiS_3$, we highlight a pseudo-2D layered perspective of $BaTiS_3$, where Ba is surrounded by three nearest $TiS_6$ octahedra, which form a triangular lattice structure. Here, when we consider the *a-b* plane asymmetry of these octahedra, the 2D layer forms an effective Kagome-type structure, as shown in **Figure 2c**. Such pseudo-2D layers are compactly stacked along the *c*-axis, exposing such surfaces in 001-oriented crystals[36]. Given the global rotational symmetry along the *c*-axis, the polarisation of Ti can be decomposed [**Extended Data Figure 5**] along and perpendicular to the *c*-axis, represented by exaggerated coloured markers.

We now summarise the polarisation textures of $BaTiS_3$ by projecting them toward the *a-b* plane. Because the *c*-gliding symmetry mirrored *a-b* plane displacements separated by *c*/2 within the unit cell, each unit cell of $BaTiS_3$ decomposes into two layers with opposing handedness: upper- and lower-unit-cell in **Figure 2c**. We discovered that the *a-b* plane dipolar displacements form alternating clockwise and anticlockwise vortices along the *a-b* plane of $BaTiS_3$. We identify three distinct patterns of vortices arising here, where each vortex comprises seven $TiS_6$ octahedra: six asymmetric $TiS_6$ surrounding a symmetric $TiS_6$. The spatial distribution of these *a-b* plane dipolar displacements exhibits three distinct vortex characteristics: anticlockwise vortex, clockwise vortex, and head-to-head-tail-to-tail antivortex, illustrated with shaded arc arrows in red, blue, and grey, respectively. Adjacent vortices have a corner shared $TiS_6$, similar to a frustrated Kagome lattice[14].

To understand the observed displacement modes, *P*3*c*1 $BaTiS_3$ is compared with the centrosymmetric structure, $P6_3/mmc$[28], as shown in **Figs. 3a-c**. Due to the triangular symmetry, the dipoles with antipolar order experience frustration from competing nearest-neighbour interactions. This frustration is relieved by forming a periodic array of two adjacent vortices with different chirality, *i.e.*, clockwise, anticlockwise, and one connecting antivortex (**Figure 3e**). Their winding numbers ($w$) are then extracted as $w = 1$, $w = 1$, and $w = -2$, respectively[49] (details in the **Supplementary Materials, Section 8**). Such *a*-*b* plane vortex triplets span 2.3 nm (corresponding to the real space dimension of $q_3$), with the edge of each vortex being 1.3 nm (corresponding to the real space dimension of $q_2$), and a thickness of 2.9 Å along the *c*-axis [**Extended Data Figure 3c**].

To evaluate the stability of the observed polarisation vortex texture, we used density functional theory (DFT) to compare the total energies of structures incorporating collinear polar displacements along the *a*-*b* plane [**Extended Data Figure 8**]. These collinear configurations are higher in energy than the vortex phase, $P3c1$ $BaTiS_3$. This rules out a scenario in which the domains are composed of collinear dipoles, with adjacent domains rotated in-plane to form a vortex-like pattern (see details in the **Supplementary Materials, Section 12**).

The emergence of the vortex triplets is dictated by the $q_3$ propagation vector; hence, tracking the corresponding satellite reflection of $q_3$ with diffraction is an effective tool for probing the evolution of the vortex order with temperature across the phase transition. We performed synchrotron X-ray diffraction measurements during the heating and cooling cycles between 230 K and 270 K in **Extended Data Figure 9c**, across the CDW transition (details see **Methods IV** and the **Supplementary Materials, Sections 7**). Given that $\frac{5}{6}\frac{\bar{1}}{6}0$ reflection, is a first-order of the satellite $q_3 = (-1/6, -1/6, 0)$ corresponding to the $BaTiS_3$ 100 reflection, and is insensitive to displacements along the *c*-axis, the vortex order transition is elucidated by tracking the evolution of $\frac{5}{6}\frac{\bar{1}}{6}0$ reflection. On one hand, the sharp transitions align more closely with a first-order-type nucleation and annihilation, as illustrated in **Extended Data Figures 9c-d**. On the other hand, the transition onset temperatures ($T_0$) during heating and cooling cycles exhibit minimal undercooling, with $T_{0,h} = 252.5$ K and $T_{0,c} = 249.0$ K. The nucleation and dissolution of $P3c1$ $BaTiS_3$ appear to have a very low energy barrier, unlike a classic nucleation-growth-driven first-order transition, where stochastic nucleation leads to significant undercooling.

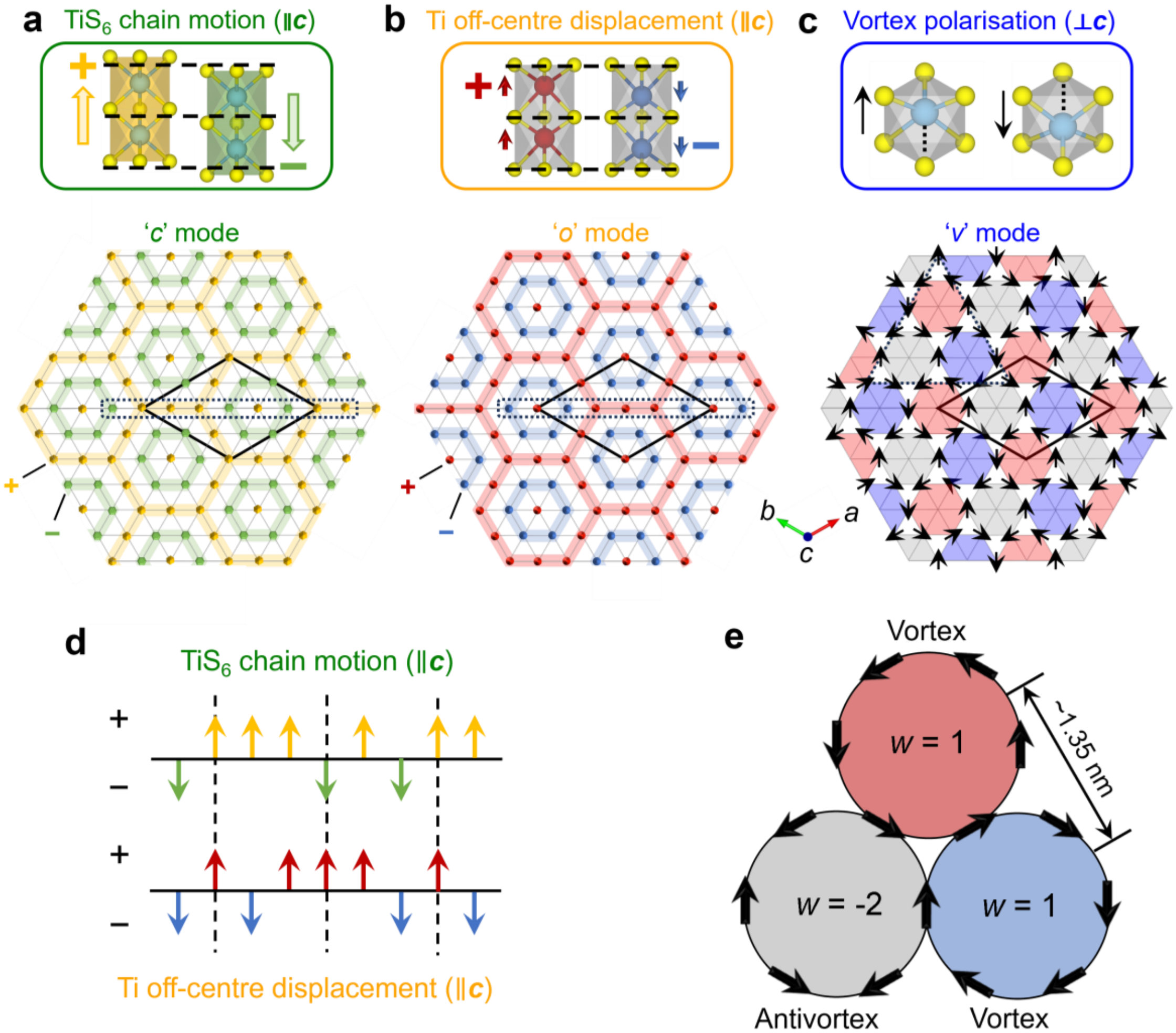

**Figure 3. Characteristic displacement modes of the CDW phase of $BaTiS_3$.** The centrosymmetric *$P6_3/mmc$* is modulated by three hybrid modes to achieve the proposed *P*3*c*1 $BaTiS_3$ and these modes are: (**a**) $TiS_6$ chain motion along the *c*-axis referred to as the '*c*' mode (chain mode); (**b**) Ti off-centring along the *c*-axis referred to as the '*o*' mode (off-centring mode); and (**c**) vortical displacements along the *a*-*b* plane, referred to as '*v*' mode (vortex mode). The topmost panels show a magnified view of the atomic displacements in each of the three modes. All the above displacement modes have the same modulation wave vector and lead to the same solid rhombus superstructure unit cell. But a phase shift in the centre of the modulation between (**a**) and (**b**) caused the point group to reduce symmetry. This is further illustrated in (**d**), breaking the (top) mirror symmetry by off-phase displacements (bottom) along the *c*-axis, within the dotted boxes in (**a)** and (**b)**. (**e**) Vortex triplet as in the dotted triangle in (**c**). Vortices and antivortices have winding numbers $w = 1$ and –2, respectively.

## Origin of the Vortex Phase

As we closely examine the collective dipolar displacements of the $P3c1$ $BaTiS_3$ from the $P6_3/mmc$ basic structure, we identified three major hybrid (not orthogonal to each other) displacement modes: antiparallel displacements of the chain motion, referred to as "*c*" mode, the off-centring of Ti along the *c*-axis, referred to as "*o*" mode, and the triplet vortex dipolar displacements in the *a-b* plane, referred to as "*v*" mode, as shown in **Figures 3a-c**. These three modes are modulated by a common set of wave vectors, $q_1$, $q_2$, and $q_3$. However, the centre of modulation for the "*c*" and "*o*" displacement modes is spatially offset by two inter-chain distances to match the symmetry of '*v*', as illustrated in **Figures 3a-c**. This leads to the breaking of mirror symmetry, illustrated in **Figure 3d**, which agrees well with the non-zero SHG with polarisation perpendicular to the *c*-axis.

To understand the origin of the polarisation topology in the CDW phase of $BaTiS_3$, we performed group-theoretical analysis and first-principles DFT calculations. We have identified a total of 10 distortion modes that connect the $P6_3/mmc$ phase to the $P3c1$ phase [**Extended Data Figure 11a**], including three primary distortion modes, $\mathrm{K}_3$, $\Lambda_3$, $\mathrm{M}_2^-$, and one secondary distortion $\Lambda_1$. Minor distortions, including $\Gamma_1^+$, $\Gamma_2^-$, $\Gamma_3^-$, $\mathrm{K}_1$, $\mathrm{M}_1^+$, $\mathrm{M}_4^+$, and $\mathrm{M}_3^-$ modes, which collectively contribute ~5% of the total amplitude, were neglected for simplicity. Accordingly, in the phonon band structure of $P6_3/mmc$ phase (**Figure 4a**), we observe flat bands with negative frequencies along the M-Γ-Λ-K direction, which suggests the presence of competing lattice instabilities with multiple symmetry-lowering displacement modes[50] (see details in the **Supplementary Materials, Section 9**). The individual soft-phonon branches lead to the hybrid displacement modes observed in **Figures 3a-c**. Specifically, the '*o*' mode and the '*c*' mode along the *c*-axis arise from the coupled distortions introduced at the K, Λ, and M points along the phonon branches centred around -5.6 THz (green colour) and -0.9 THz (blue colour), respectively. The '*v*' mode primarily arises from the coupled distortions introduced at the Λ and M points along the phonon branch centred around -3.7 and -3.2 THz (brown and purple colour). Thus, the phonon instabilities have all the ingredients to stabilise the *P*3*c*1 lattice by symmetry lowering. Each of these instabilities is an interplay between long-range interactions (electrostatic and strain) and short-range gradient energies in the bulk $BaTiS_3$, and they collectively contribute to the realised intricate polarisation texture in $BaTiS_3$.

To assess the energetic stability, we employed the '*c*', '*o*', '*v*' displacement modes as hybrid order parameters within a Landau phase transition model: $Q_c$, $Q_o$, and $Q_v$, corresponding to the "*c*", "*o*", and "*v*" modes in **Figures 3a-c**, respectively. The resulting energy landscape as a function of their amplitudes is illustrated in **Figure 4b** (using the equation given in the **Supplementary Materials, Section 10**). Although each mode individually reduces the energy from that of the centrosymmetric $P6_3/mmc$ space group, the *P*3*c*1 space group incorporates all three modes achieves an overall lower energy state. This suggests that a coupling between the hybrid order parameters leads to an appreciable decrease in energy. We tracked the energy of the coupling terms in the Landau function (**Figure 4c**). We find that the coupling between the $TiS_6$ chain ($Q_c$) and Ti vortices ($Q_v$), as described by bi-quadratic and linear-cubic terms in the Landau function [**Extended Data Table 6**], is the most stabilising term when they act along the same direction (denoted as '$c^+ \oplus v^+$'). We attribute the energy lowering due to their coupling to an adjustment of the coordination environments around Ba and Ti cations, as quantified by an analysis of their bond-valence sums[51] (see details in the **Supplementary Materials, Section 11**). We show a comparison of the energy contributions of the individual displacement modes and their various coupled forms towards the stabilisation of the *P*3*c*1 and $P6_3cm$ phases in **Figure 4d**. In the $P6_3cm$ phase, individual modes, such as $\Gamma_2^-$ and $K_3$, dominate the total energy. In the *P*3*c*1 phase, the two modes related to Ti off-centring ($Q_o$) and $TiS_6$ chain motions ($Q_c$) along the *c*-axis exhibit large energy gains, while the vortex ($Q_v$) mode leads to negligible energy gain. However, the coupling between $Q_c$ and $Q_v$ significantly contributes to the stabilisation, eventually favouring the *P*3*c*1 over the room-temperature $P6_3cm$. Thus, our analysis shows that an intricate coupling with the out-of-plane modes provides additional freedom to stabilise the metastable polar vortex structure observed experimentally in the *P*3*c*1 phase of $BaTiS_3$.

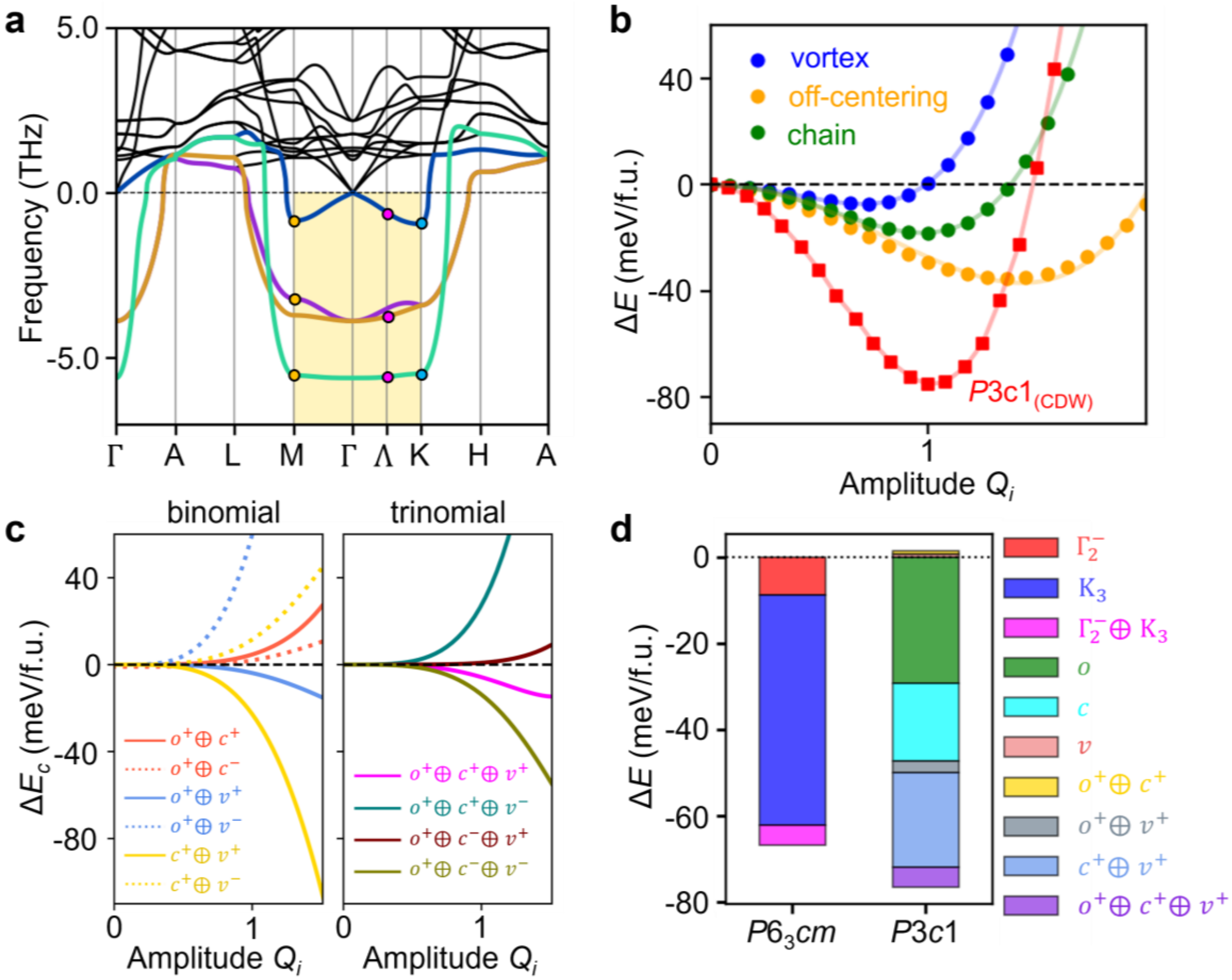

**Figure 4. Lattice instabilities that lead to the vortex triplets in the CDW phase of $BaTiS_3$.** (**a**) Phonon band structure of the centrosymmetric $P6_3/mmc$ phase. Phonon branches having symmetry-lowering soft modes are indicated by coloured lines with the green branch resulting in Ti off-centring along the *c*-axis, the brown and purple branches resulting in the *a*-*b* plane Ti displacements, and the dark blue branch resulting in $TiS_6$ chain motion along the *c*-axis. The selected high-symmetry points in the Brillouin zone are Γ (0,0,0), M ($\frac{1}{2}$,0,0), K ($\frac{1}{3}$, $\frac{1}{3}$, 0), Λ($\frac{1}{6}$, $\frac{1}{6}$,0), A (0, 0, $\frac{1}{2}$), L ($\frac{1}{2}$, 0, $\frac{1}{2}$), and H ($\frac{1}{3}$, $\frac{1}{3}$, $\frac{1}{2}$). Coloured dots highlight

representative symmetry-lowering displacement modes along the high-symmetry directions in the reciprocal space: $M_2^-$ in yellow, $\Lambda_3$ in pink, and $K_3$ in azure. (**b**) The potential energy of the three primary modes is shown in **Figure 3**, and that of the $P3c1$ phase as a function of displacement amplitude ($Q$) starting from the $P6_3/mmc$ phase. A strong coupling between $Q_o$, $Q_c$, and $Q_v$ displacement modes lowers the energy of $P3c1$-$BaTiS_3$ compared to the phases obtained by freezing the three modes individually. (**c**) Evolution of energy for the coupling terms ($\Delta E_c$) in the Landau model as a function of the amplitude of different combinations of the three primary modes $Q_o$, $Q_c$, and $Q_v$. The left panel shows the binomial coupling terms ($Q_i Q_j \neq 0$, $Q_k = 0$). Solid and dotted lines represent the cases where $Q_i Q_j > 0$ and $Q_i Q_j < 0$, respectively. The right panel shows the trinomial coupling terms ($Q_i Q_j Q_k \neq 0$). '+' and '-' refer to the direction of each order parameter, where the displacement patterns along the '+' direction are shown in **Figures 3a-c**. (**d**) Energy contributions of the individual and coupled terms for the experimentally observed $P6_3cm$ and $P3c1$ phases.

## Outlook

We have reported the existence of an intricate, but long-range ordered, polarisation vortex network composed of sub-unit-cell vortex-vortex-antivortex triplets of $TiS_6$ dipolar displacements in a dilute *d*-orbital bulk semiconductor, $BaTiS_3$. The three crystallographically distinct $TiS_6$ octahedral sites observed by synchrotron single-crystal X-ray diffraction (SC-XRD) are further corroborated by $^{47/49}$Ti solid-state nuclear magnetic resonance (SSNMR) spectroscopy. We then demonstrated that this complex topology of non-collinear dipoles is stabilised by *intrinsic* coupled lattice instabilities using first-principles calculations and phenomenological modelling. The coincidence of the CDW-like electronic order, antiparallel dipole order, and the triplet polarisation vortex textures in zero-filling semiconductors is anticipated to fuel the study of anomalous polar interactions in effective Kagome-type triangular lattices in quasi-1D materials beyond $BaTiS_3$, inspiring the discovery of vibrant polarisation textures.

The non-trivial atomic-scale polarisation topology realised here holds importance in understanding the size limit of dipolar topological structures and the intrinsic mechanisms controlling their stability. Recent reports on sub-terahertz (THz) dynamics in nanoscale polar vortices suggest the potential for strong light-matter interactions in the THz and far-infrared regime.[52] Further, recent reports on the unique interaction of polar materials with twisted light[53] and the realisation of light-induced chiral order[54] present a unique opportunity to explore and realise dynamic, complex atomic-scale polarisation textures in materials such as $BaTiS_3$.

# Methods

## I. Materials Synthesis

Single crystals of $BaTiS_3$ were grown by chemical vapour transport with iodine as a transporting agent. Starting materials, barium sulphide powder (Sigma-Aldrich, 99.9%), titanium powder (Alfa Aesar, 99.9%), sulphur pieces (Alfa Aesar, 99.999%), and iodine pieces (Alfa Aesar 99.99%) were stored and handled in a nitrogen-filled glove box. Stoichiometric quantities of precursor powders with a total weight of 1.0 g were mixed and loaded into a quartz tube of diameter 19 mm along with ~0.75 $mg \cdot cm^{-3}$ iodine inside the glove box. The tube was capped with ultra-torr fittings and a bonnet needle valve to avoid exposure to the air. The tube was then evacuated and sealed using a blowtorch. The sealed tube was about 12 cm in length and was loaded and heated to 1055 °C at 100 °C/h, held for 100 h, and then slowly cooled down to 950 °C at 10 °C/h before being cooled down with the furnace.

## II. Temperature-Dependent and High-Resolution XRD

Temperature-dependent XRD was carried out at the 7-ID-C beamline of the Advanced Photon Source. A $BaTiS_3$ single crystal was mounted on a silicon substrate with silver epoxy as adhesive, which was also used to load the substrate on a closed-cycle cryostat-cooled six-circle diffractometer. The diffractometer was equipped with a 2D detector (Pilatus 100 K). Monochromatic hard X-ray radiation with a wavelength of 1.239 Å was focused to 50 µm full-width-half-maximum (FWHM) at the sample position.

Temperature is controlled by the Lakeshore 340 temperature controller and is swept between 230 K and 270 K with a step size of 1 K during heating and 2 K during cooling of $BaTiS_3$. Due to the low X-ray incident angle, height changes in the sample stage as a function of temperature result in major misalignment. A *z*-scan about the average sample height was utilised to offset the sample-stage displacements. The rocking curve, however, is wide with full width at half maximum (FWHM) ~ 0.5° and translates within ±0.05° throughout the heat/cooling cycles. the pixels of the scattering images are binned over $2\theta$ and integrated to get the intensity.

High-resolution XRD measurements were performed at the 7-ID-C beamline of the Advanced Photon Source after the APS-U upgrade, using a focused beam of ~20 μm and an Eiger2X 500k detector. Reciprocal space mapping is performed by conducting a rocking curve scan around the peak of the 100 reflection. 2D images are pixelated and transformed to reciprocal space, then integrated into the reciprocal space map in RSMap3D. The peak profile along the *a*-axis and the *c*-axis is then extracted through the centre of the peak and fitted as Pseudo-Voigt profiles to get the FWHM. Correlation length ξ is the inverse of FWHM.

## III. Single Crystal Diffraction

Laboratory X-ray single-crystal diffraction was carried out on the Rigaku XtaLAB diffractometer at the University of Southern California. The diffractometer is equipped with a Rigaku (Mo Kα) rotating-anode X-ray tube and a hybrid pixel array detector. Crystals cleaved into 50 μm × 20 μm × 10 μm pieces were mounted on Mi-TeGen Kapton loops and placed in a nitrogen cold stream. Diffraction data were collected using a monochromatic source with a wavelength of 0.71073 Å.

High-resolution synchrotron single-crystal diffractions were carried out at beamline 12.2.1 at the Advanced Light Source, Lawrence Berkeley National Laboratory. The crystal was mounted on Mi-TeGen Kapton loops and placed in a nitrogen cold stream on the goniometer head of a Bruker D8 diffractometer, which is equipped with a PHOTONII CPADS detector operating in shutterless mode. Diffraction data were collected using synchrotron radiation monochromated with a wavelength of 0.72880 Å, using a silicon (111) crystal. Weak reflections are repeated with stronger (approximately 50×) incident synchrotron radiation. Unit cell determination, integration, and scaling are then carried out in Bruker APEX 3, during which process the precession map was generated with a thickness of 0.1 (in reciprocal space indices) under the refined unit cell.

Crystal structure refinement and electron density map analysis were done in ShelXle[38,55]. All possible space groups and pseudo-merohedral twins are tested to find the best-matching crystal structure.

## IV. Rotation Anisotropy Second Harmonic Generation (RA-SHG)

Second Harmonic Generation (SHG) measurements were carried out using an ultrafast laser source with a central wavelength of 800 nm, a repetition rate of 200 kHz, and a pulse duration of 80 fs. Rotation anisotropy (RA) SHG experiments were performed by measuring the SHG intensity as a function of light polarisation rotation with respect to the in-plane crystal axes, with the incident fundamental and reflected SHG polarisations set either parallel or perpendicular to each other. In this study, due to the small domain size associated with the charge density wave states, the incident laser beam was tightly focused to a spot size of approximately 2.5 μm using a 20× objective lens with a numerical aperture (N.A.) of 0.42. Consequently, the RA SHG measurements were conducted in a normal-incidence geometry.

## V. Density Functional Theory (DFT) Analysis of EFG Tensors and $TiS_6$ Local Symmetries

Quantum chemical calculations of titanium EFG tensors were performed within the framework of plane-wave DFT as implemented in the CASTEP module of Materials Studio 2020[56–58]. These calculations were performed on model structures obtained from previous diffraction studies. Calculations employed the PBE functional with Wu-Cohen exchange, which is a generalised-gradient approximation method optimised for calculations on solids[59,60]. Core-valence interactions were modelled with ultrasoft pseudopotentials[61] generated on the fly with relativistic effects incorporated through the ZORA pseudopotential formalism[62]. The wavefunction was expanded in a basis described by a plane-wave cutoff energy of 1000 eV. Integrals over the Brillouin zone were sampled using a Monkhorst-Pack[63] grid with a $k$-point spacing of 0.06 $Å^{-1}$. SCF convergence was achieved through the All-Bands/EDFT minimiser using a threshold of $5 \times 10^{-8}$ eV $atom^{-1}$.

## VI. Solid-State Nuclear Magnetic Resonance (SSNMR)

$^{47/49}Ti$ experiments were performed using a Bruker NEO console with an Oxford 18.8 T ($\nu_0(^1H)$ = 800 MHz) magnet operating at $\nu_0(^{49}Ti)$ = 45.112 MHz.

A 5 mm HX static probe developed by PhoenixNMR was used for all experiments, with samples packed into cylindrical 5.0 mm o.d. polychlorotrifluoroethylene sample containers designed at MagLab, which reduce $^1H$ background signals and allow storage of air-sensitive samples. Variable-temperature experiments were performed using a temperature regulation unit developed by PhoenixNMR. All SSNMR spectra were acquired under static conditions (*i.e.*, stationary samples).

All $^{47/49}Ti$ SSNMR spectra were acquired using the WURST-CPMG pulse sequence[64,65], which utilises frequency-swept WURST (Wideband Uniform Rate Smooth Truncation) pulses for broadband excitation and refocusing[66]. Due to favourable effective transverse relaxation times, $T_2^{eff}$, for $^{47}Ti$ and $^{49}Ti$ in $BaTiS_3$, WURST-CPMG experiments permitted the acquisition of numerous echoes, which aids in enhancing the signal-to-noise ratio. The sweep widths of all WURST pulses were chosen such that the sweep width was greater than twice the breadth of the $^{47/49}Ti$ powder pattern in kHz to ensure uniform excitation.

Pulse width calibrations and chemical shift referencing were carried out using solid strontium titanate, $SrTiO_3$(s). The chemical shift was referenced to neat $TiCl_4$(liq) with $\delta_{iso}$ = 0 ppm, using $SrTiO_3$(s) as a secondary standard with $\delta_{iso}$ = - 863 ppm[67].

Details on experimental conditions and acquisition parameters for all SSNMR experiments are summarised in **Tables S1-S3**. Data were acquired and processed using the TopSpin 4.1.3 software. Numerical simulations and spectral fits were performed using the ssNake software package[68].

## VII. First Principles Calculations of the Energetic Stability and the Polarisations

We performed DFT calculations using the Vienna Ab initio Simulation Package (VASP[69]). We used projector augmented-wave (PAW) potentials[70] and the generalised gradient approximation within the Perdew-Burke-Ernzerhof (GGA-PBE[71]) parameterisation to describe the electron-ion and the electronic exchange-correlation interactions, respectively. The PAW potentials explicitly included $5s^2 5p^6 6s^2$ for Ba, $3p^6 3d^2 4s^2$ for Ti, and $3s^2 3p^4$ for S as valence electrons. The initial structures of various phases of $BaTiS_3$ were sourced from the literature[31] and subsequently fully optimised using DFT. An energy cut-off of 650 eV was used for the expansion of the plane waves. The convergence criterion for the SCF energy was set to $10^{-8}$ eV. All structures were fully relaxed until the total forces on each atom were smaller than 0.01 eV/Å. The Brillouin zone was sampled using a Γ-centred Monkhorst-Pack[63] $k$-points mesh, with a maximum spacing of 0.05 $Å^{-1}$ for geometry optimisations and 0.03 $Å^{-1}$ for static calculations, respectively. The phonon dispersion for the $P6_3/mmc$ phase was calculated with the *phonopy*[72] package via the finite-displacement method using a 2 × 2 × 2 supercell.

The local polarisation shown in **Extended Data Figure 5** was calculated using the Berry phase approach[73], referenced to the centrosymmetric $P6_3/mmc$ structure. To isolate the contributions of local distortion modes, we froze phonon modes at the Γ point. For out-of-plane polarisation ($P_{\parallel}$) along the $c$-axis, we froze the $\Gamma_2^-$ mode. For the in-plane polarisation ($P_{\perp}$) along the $a$–$b$ plane, the $\Gamma_5^-$ mode was applied, but only of the two layers in the unit cell to suppress alternating layer displacement that would otherwise cancel the net polarisation. We applied a Hubbard $U$ of 3 eV for Ti atoms, lifting any metallic states in the intermediate images along the polarisation path.

## Methods References

## Acknowledgements

We thank Nuh Gedik, Abhay Pasupathy, Han Wang, Li Yang, Chong Wang, and Di Xiao for the discussions. This work was primarily supported by the Army Research Office (ARO) under an ARO MURI program with award number W911NF-21-1-0327. The modelling and experimental analysis aspects of the project were partially supported by the US National Science Foundation (NSF) through award numbers DMR-2122070, DMR-2122071, and DMR-2145797. Laboratory single-crystal diffraction instrumentation was supported by NSF award number CHE-2018740. The crystal growth instrumentations were partially supported by an Office of Naval Research grant N00014-23-1-2818. B.Z. and H.W. were supported by the U.S. Department of Energy (DOE), Office of Science (SC), Basic Energy Sciences, Materials Sciences and Engineering Division. The NMR measurements were supported by the U.S. Department of Energy, Office of Science, Basic Energy Sciences, under Award DE-SC0022310. R.S., S.T., F.M.-V., and R.W.S. are also grateful for support from The Florida State University and the National High Magnetic Field Laboratory, which is funded by the National Science Foundation Cooperative Agreement (NSF/DMR-2128556), and the State of Florida. L. Z. acknowledges the support from the U.S. Department of Energy (DOE), Office of Science, Basic Energy Science (BES), under award No. DE-SC0024145 (for acquiring and analysing RA SHG data) and the National Science Foundation through the Materials Research Science and Engineering Center at the University of Michigan, Award No. DMR-2309029 (SHG set up development and data acquisition). This research used the synchrotron resources of the Advanced Light Source, which is a DOE Office of Science User Facility under contract no. DE-AC02-05CH11231; and the Advanced Photon Source, a DOE Office of Science user facility at Argonne National Laboratory, and is based on research supported by the U.S. DOE Office of Science-Basic Energy Sciences, under Contract No. DE-AC02-06CH11357. This work used computational resources through allocation DMR160007 from the Advanced Cyberinfrastructure Coordination Ecosystem: Services & Support (ACCESS) program, which is supported by NSF grants #2138259, #2138286, #2138307, #2137603, and #2138296.

## Fundings

ARO MURI program award number W911NF-21-1-0327.

National Science Foundation award numbers CHE-2018740, DMR-2122070, DMR-2122071, DMR-2145797, DMR-2309029, and ACCESS program (DMR160007) grants # 2138259, #2138286, #2138307, #2137603, and #2138296.

Office of Naval Research award number N00014-23-1-2818.

U.S. Department of Energy, Office of Science user facility under Contract No. DE-AC02-05CH11231, and No. DE-AC02-06CH11357.

U.S. Department of Energy, Office of Science, Basic Energy Science (BES), under award No. DE-SC0024145, and DE-SC0022310.

National Science Foundation Cooperative Agreement (NSF/DMR-2128556) (NHMFL)

State of Florida (NHMFL)

## Author contributions

Crystal Growth: B.Z., Z.D., Q.Z., H.C., and J.R.

Single Crystal Diffraction: B.Z., S.S., H.C., K.Y., N.S., J.R., and S.T.

Second Harmonic Generation: W.Z., C.W., and L.Z.

Solid State Nuclear Magnetic Resonance: R.B.S., S.T., F.M.-V., and R.W.S.

Landau Model: G.Y.J., M.J.S., and R.M.

Density Functional Theory: G.Y.J, G.R., M.J.S., S.T.H., and R.M.

Temperature Resolved High Resolution X-ray Diffraction: B.Z., S.S., J.R., D.W., and H.W.

High resolution XRD data analysis: B.Z., C.W., and J.R.

## Competing Interests

The authors declare no competing interests.

## Data and materials availability

The experimental data used in this study are available from the corresponding authors and at Zenodo at: 10.5281/zenodo.12171246

# Extended Data for

## Emergent Atomic Scale Polarisation Vortices in $BaTiS_3$

Boyang Zhao[1,2]†, Gwan Yeong Jung[3]†, Shantanu Singh[1], Robert B. Smith[4,5], Huandong Chen[1], Guodong Ren[6], Chuangtang Wang[7], Sara Termos[4,5], Sean T. Holmes[4,5], Frederic Mentink-Vigier[4,5], Weizhe Zhang[7], Zhengyu Du[1], Claire Wu[1], M. J. Swamynadhan[3], Qinai Zhao[1], Kevin Ye[1], Donald A. Walko[8], Nicholas S. Settineri[9], Simon J. Teat[9], Liuyan Zhao[7], Robert W. Schurko[4,5], Haidan Wen[2,8], Rohan Mishra[3,6,*], and Jayakanth Ravichandran[1,10,11,*]

[1]*Mork Family Department of Chemical Engineering and Materials Science, University of Southern California, Los Angeles, CA 90089, USA*

[2]*Materials Science Division, Argonne National Laboratory, Lemont, IL 60439, USA*

[3]*Department of Mechanical Engineering & Materials Science, Washington University in St. Louis, St. Louis, MO 63130, USA*

[4]*Department of Chemistry and Biochemistry, Florida State University, Tallahassee, FL 32306, United States*

[5]*National High Magnetic Field Laboratory, Tallahassee, Florida 32310, United States*

[6]*Institute of Materials Science & Engineering, Washington University in St. Louis, St. Louis, MO 63130, USA*

[7]*Department of Physics, University of Michigan, Ann Arbor, MI 48019, USA*

[8]*Advanced Photon Source Argonne National Laboratory, Lemont, IL 60439, USA*

[9]*Advanced Light Source, Lawrence Berkeley National Laboratory; Berkeley, CA 94720, USA*

[10]*Ming Hsieh Department of Electrical Engineering, University of Southern California; Los Angeles, CA 90089, USA*

[11]*Core Center of Excellence in Nano Imaging, University of Southern California; Los Angeles, CA 90089, USA*

† These authors contributed equally: Boyang Zhao, Gwan Yeong Jung

* Corresponding author. Email: j.ravichandran@usc.edu, rmishra@wustl.edu

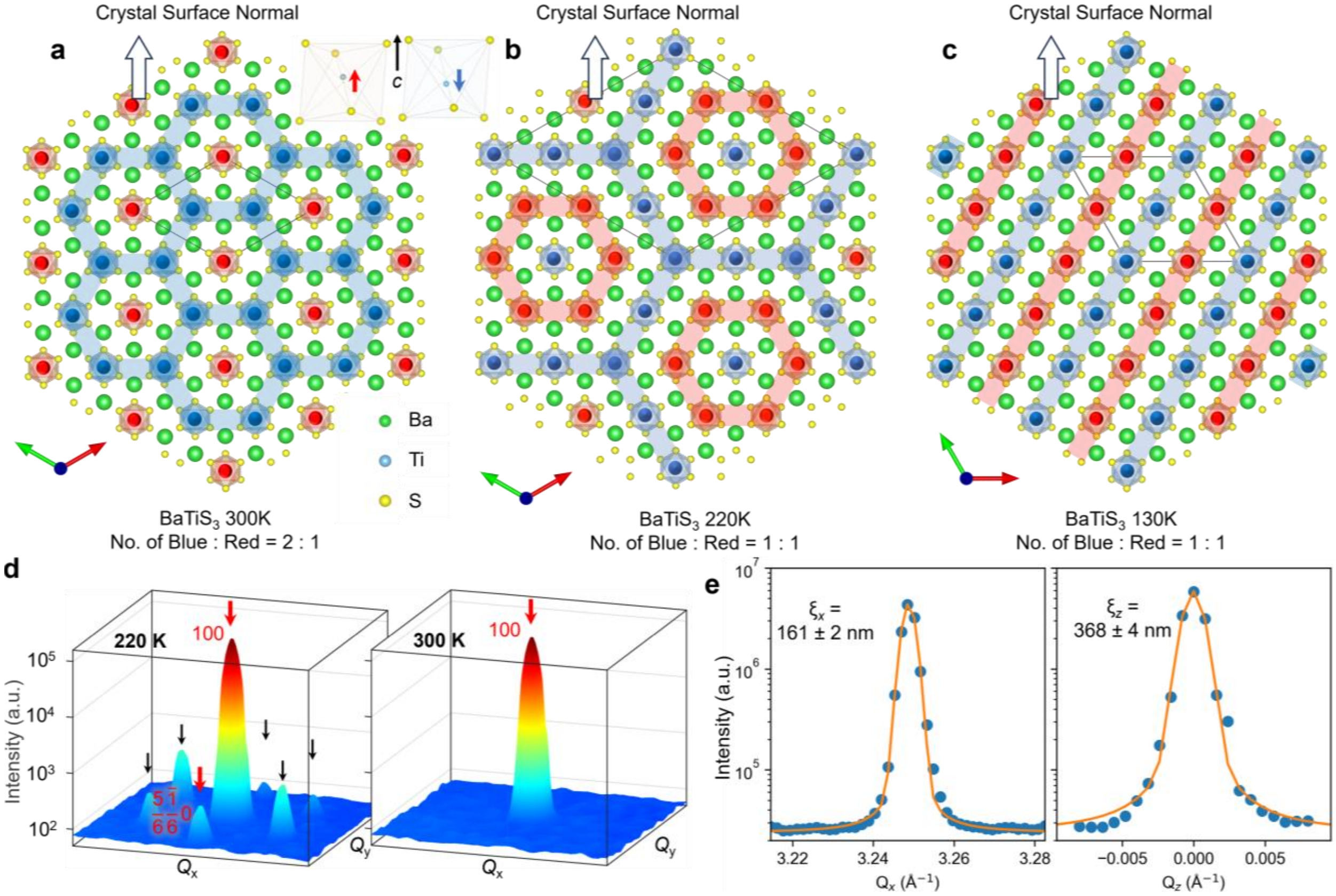

**Extended Data Figure 1. Dipole order of dipolar $TiS_6$ octahedra along the *c*-axis in $BaTiS_3$.** Ti atoms are displacing away from the centre of the $S_6$ octahedra (visualised in the inset in (**a**)) along the same orientation along the *c*-axis for each $TiS_6$ chain. We thus project the $BaTiS_3$ structure along the *c*-axis and colour each $TiS_6$ chain as Ti out-of-centre displacements pointing down (blue) or up (red) along the *c*-axis. The resulting dipolar $TiS_6$ octahedra show different "ferroic" ordering, which is (**a**) 300 K hexagonal ferrielectric order of $\sqrt{3} \times \sqrt{3} \times 1$ unit cell, $P6_3cm$; (**b**) 220 K CDW-like hexagonal antiferroelectric order of $2\sqrt{3} \times 2\sqrt{3} \times 1$ unit cell, $P3c1$; and (**c**) 130 K lamella monoclinic antiferroelectric order of $2 \times 2 \times 1$ unit cell, $P2_1$. (**d**) The reciprocal space map (RSM) of the $BaTiS_3$ crystals shows analogous peak profiles between the satellite peak and the main peaks. (**e**) Correlation length of $BaTiS_3$ at 200 K, extracted from the RSM analysis of the 100 reflection obtained using high-resolution XRD. The correlation length within the *a*-*b* plane ($\xi_x \sim 161 \pm 2$ nm) is marginally smaller than along the *c*-axis ($\xi_z \sim 368 \pm 4$ nm), but well over the superstructure unit cell dimension of ~ 2.3 nm.

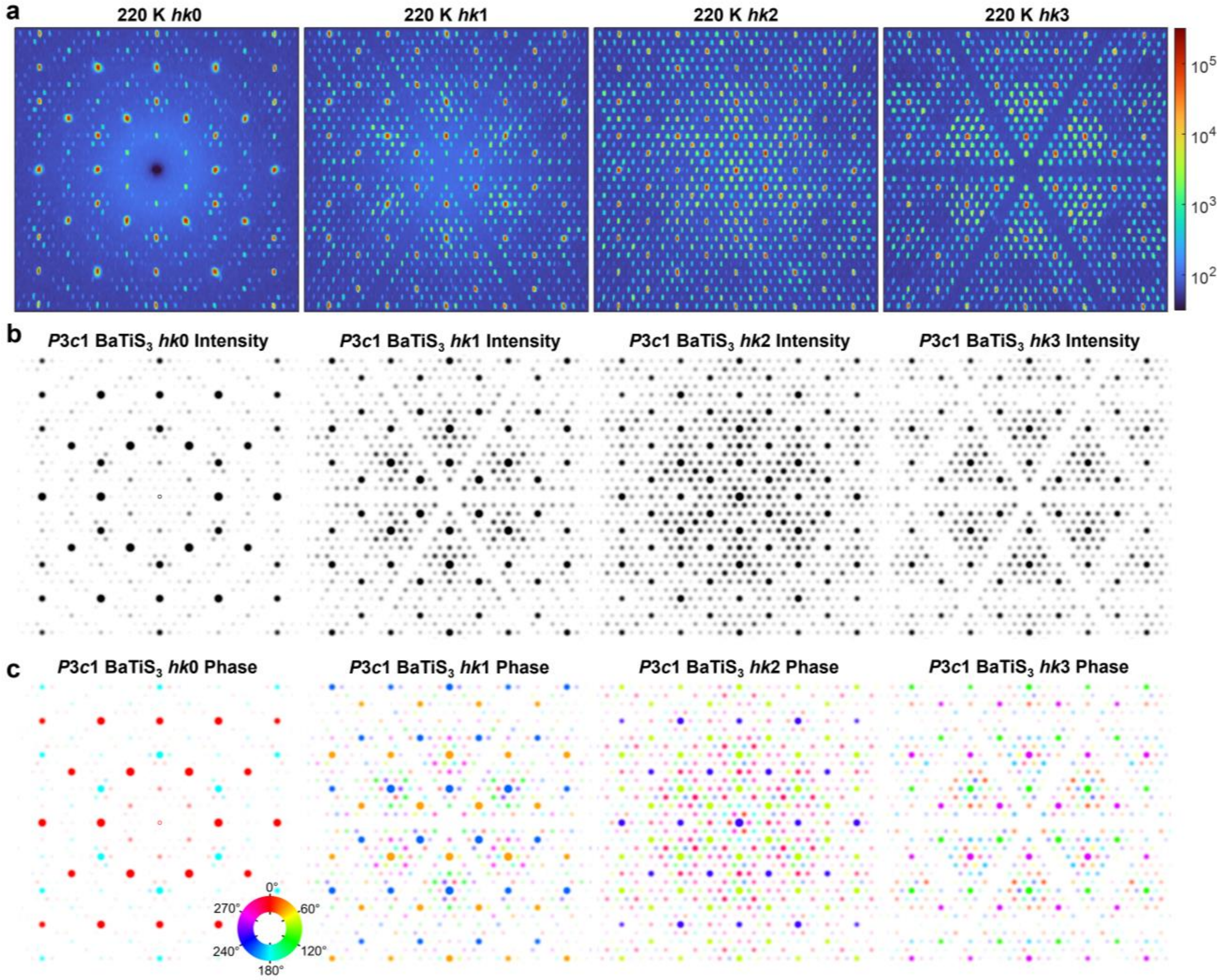

**Extended Data Figure 2. Extended Reciprocal Space map of 220K $BaTiS_3$. (a)** The *hkl* ($l$ = 0, 1, 2, 3, …) SC-XRD precession maps, integrated from the SC-XRD collected on $BaTiS_3$ crystal at 220 K. Diffraction symmetry (Laue symmetry $\bar{3}m$) is achieved by analysing the overall intensity profile. Friedel pairs (*i.e.*, *hkl* and $\bar{h}\bar{k}\bar{l}$) reveal the existence of inversion operations either through an inversion symmetry (excluded during refinement) or an inversion twin. The intensity extinctions at *hhl* ($l$ = odd) are further used to examine the *c*-glide crystalline symmetry towards *P*3*c*1. (**b-c**) The *hkl* ($l$ = 0, 1, 2, 3, …) precession photos of the simulated XRD structure factors of the *P*3*c*1 $BaTiS_3$. Peaks are coloured according to their (**b**) intensities and (**c**) phase.

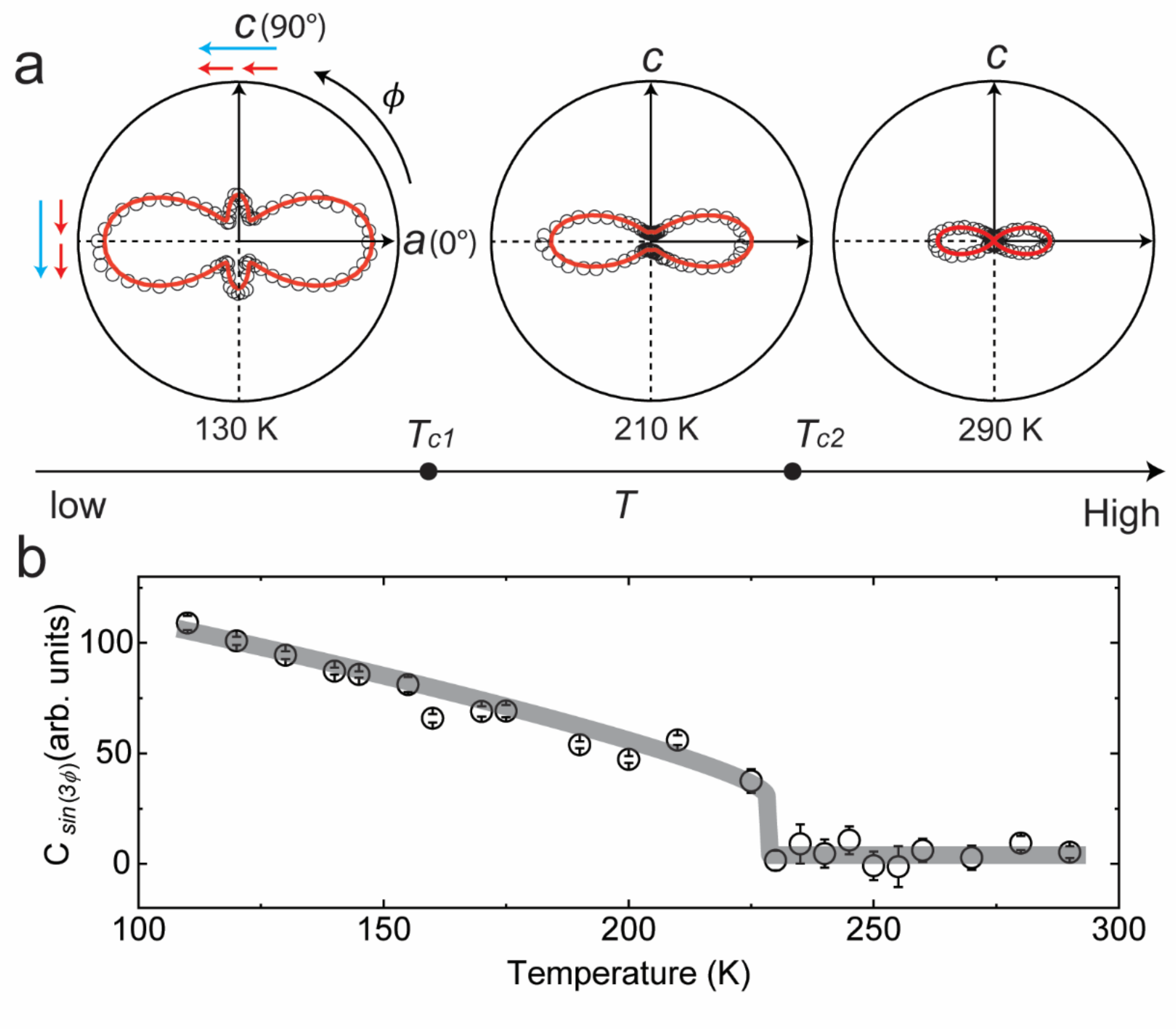

**Extended Data Figure 3. SHG Results Revealing Mirror Symmetry Breaking Across CDW Transition.** (**a**) RA-SHG patterns across transition temperatures. High temperature (290 K) features a mirror symmetry with zero SHG intensity when SHG polarisation is normal to the mirror plane. When $T < T_{c2}$, the mirror symmetry is broken with a finite SHG background (210 K) and small lobes (130 K). Open circles represent experimental data, and solid red curves represent the best-fit results with the domain-average functional form for data at 130 K and 210 K, and the single-domain functional form for data at 290 K. Red and blue arrows indicate the polarisation of fundamental and SHG light at the corresponding angle. **b**, Fitting $\mathrm{C}_{\sin(3\phi)}$ parameter across transition temperatures, its onset behaviour indicates the breaking of mirror symmetry and the transition to the CDW state. The black line is a guide for the eye.

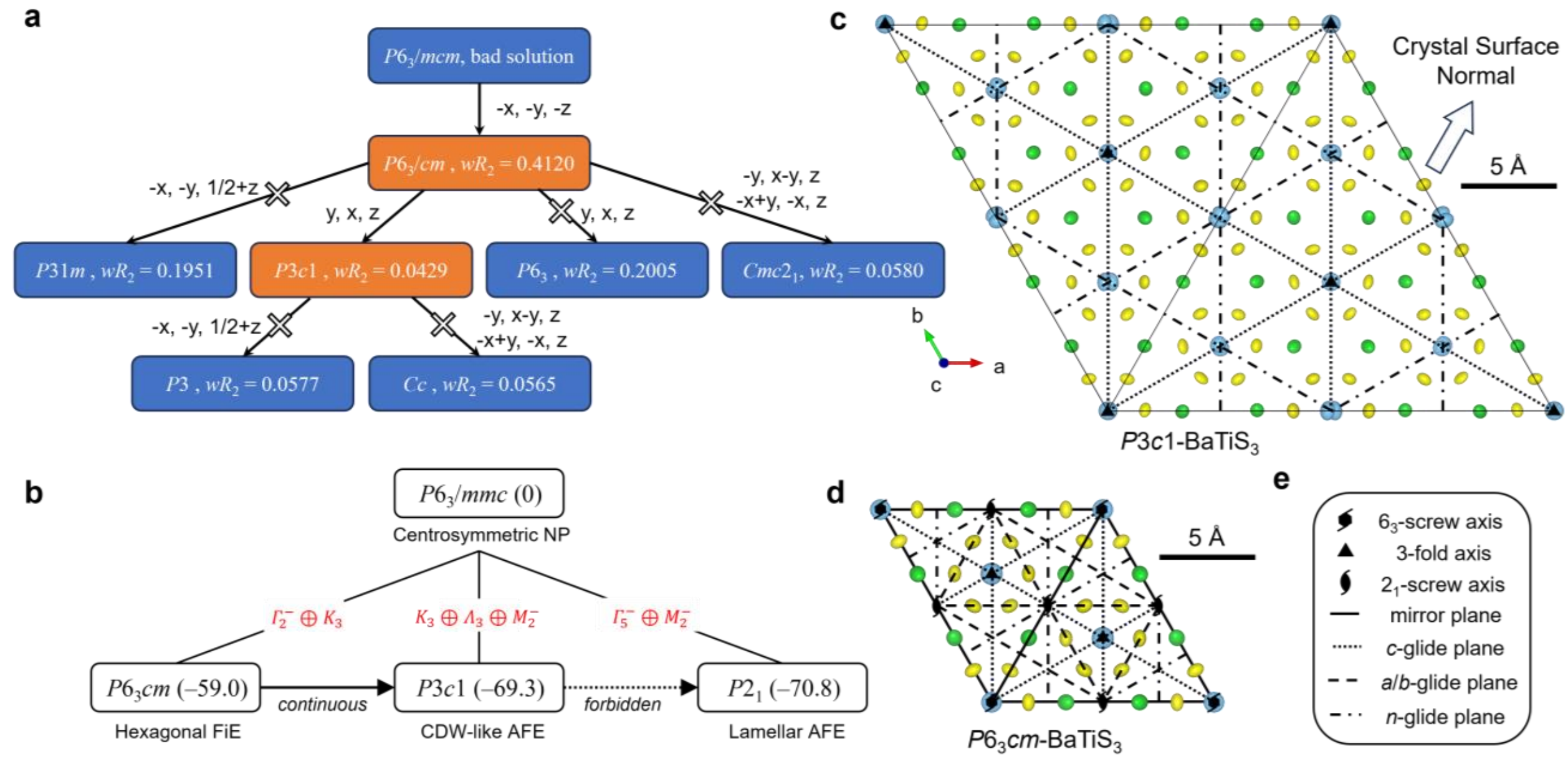

**Extended Data Figure 4. Potential symmetry solutions for the phases during the phase transition of $BaTiS_3$.** (a) The symmetry determination flowchart represents the refined structure solutions for the $BaTiS_3$ diffraction data collected at 220 K. The flowchart is presented as a group-subgroup relationship plot. Each solution is refined to match the SC-XRD, and the refinement residual, $wR_2$, reflects how well the described solutions, $|F_{hkl}|^2$, fits the experimental observations, $I_{hkl}$. Symmetry operations that are broken by the subgroup are noted beside the arrows, and they are used as a twinning operation in the low-symmetry space groups. A lower residual in the low symmetry group is thus a better solution, while a higher or comparable residual makes the high symmetry space group a better solution. (**b**) Group-subgroup relationship for the observed phase transitions in $BaTiS_3$, starting from the parent, centrosymmetric non-polar (NP) $P6_3/mmc$ phase to the low-symmetry phases observed in experiments, which are $P6_3cm$ at 300 K, $P3c1$ at 220 K, and $P2_1$ at 130 K[31]. Synchrotron X-ray refinements show three distinct antipolar ordering patterns along the $c$-axis in these low-symmetry phases: a hexagonal ferrielectric (FiE) order in $P6_3cm$, a CDW-like antiferroelectric (AFE) order in $P3c1$, and a lamellar monoclinic antiferroelectric order in $P2_1$; see **Extended Data Figure 1**. The displacement modes are labelled using the $P6_3/mmc$ as the reference structure. The numbers in parentheses denote the relative energy in units of meV/f.u.. (**c-e**) Symmetry elements of $BaTiS_3$ overlaid with an $a$-$b$ plane projection of (**c**) $P3c1$-$BaTiS_3$ and (**d**) $P6_3cm$-$BaTiS_3$[32].

Symmetry elements used are summarised in (**e**). Note that (**c**) and (**d**) have the same atomic density but differ in unit cell size.

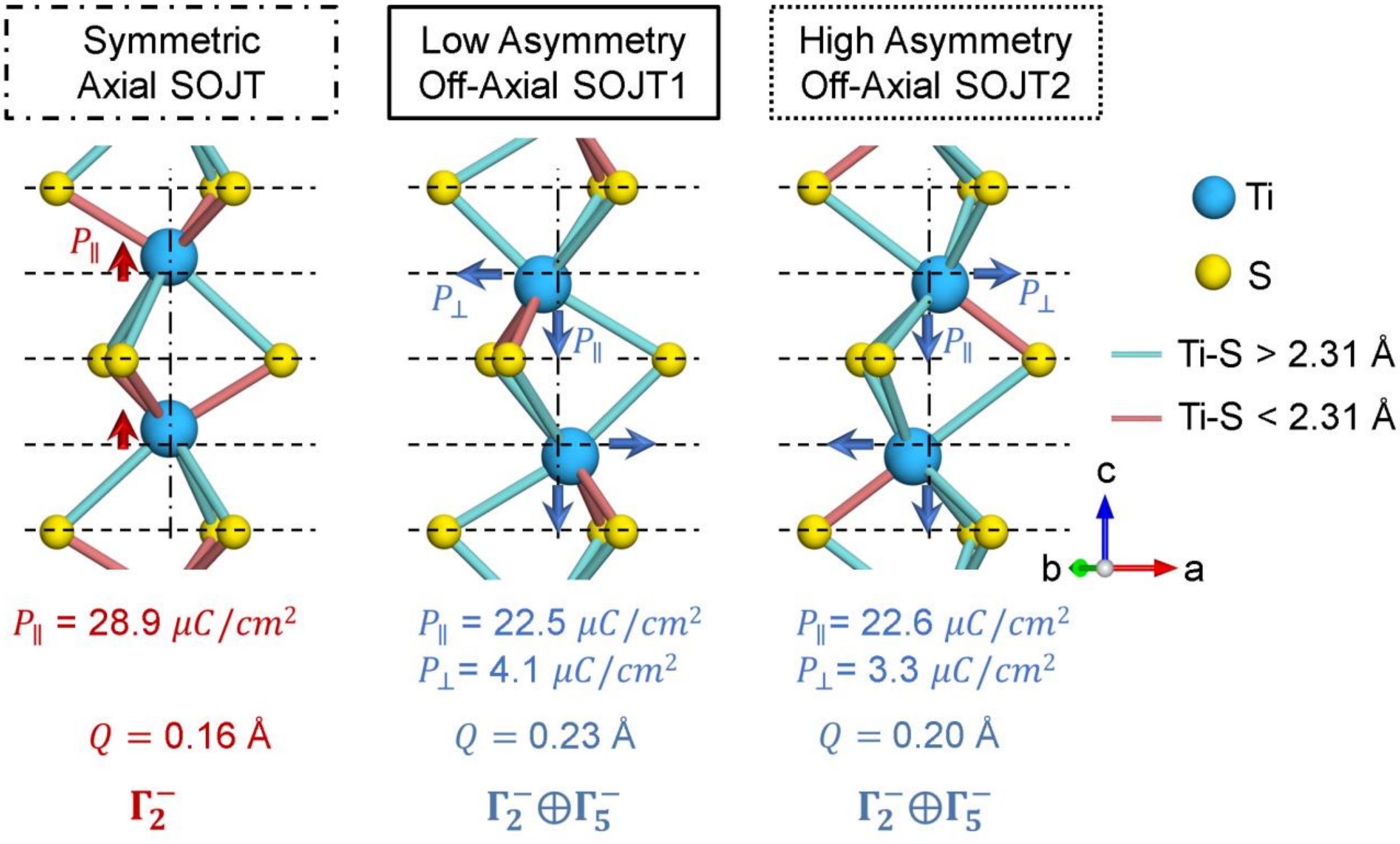

**Extended Data Figure 5. Local displacive distortions of $TiS_6$ octahedra projected onto the *a*-*c* plane.** Three $TiS_6$ motifs constitute the $P3c1$ phase: (left) axially symmetric Ti- off-centre displacements, (middle) low axial asymmetry, and (right) high axial asymmetry. Along with axial Ti off-centre displacements associated with the $\Gamma_2^-$ mode, off-axial distortions of $TiS_6$ octahedra arising from the coupled $\Gamma_2^- \oplus \Gamma_5^-$ mode give rise to zigzag-like polarisation along the chains. The Ti out-of-centre displacements are marked with coloured arrows. Polarisation components ($P_{\parallel}$ and $P_{\perp}$) obtained from Berry phase calculations are indicated below each motif.

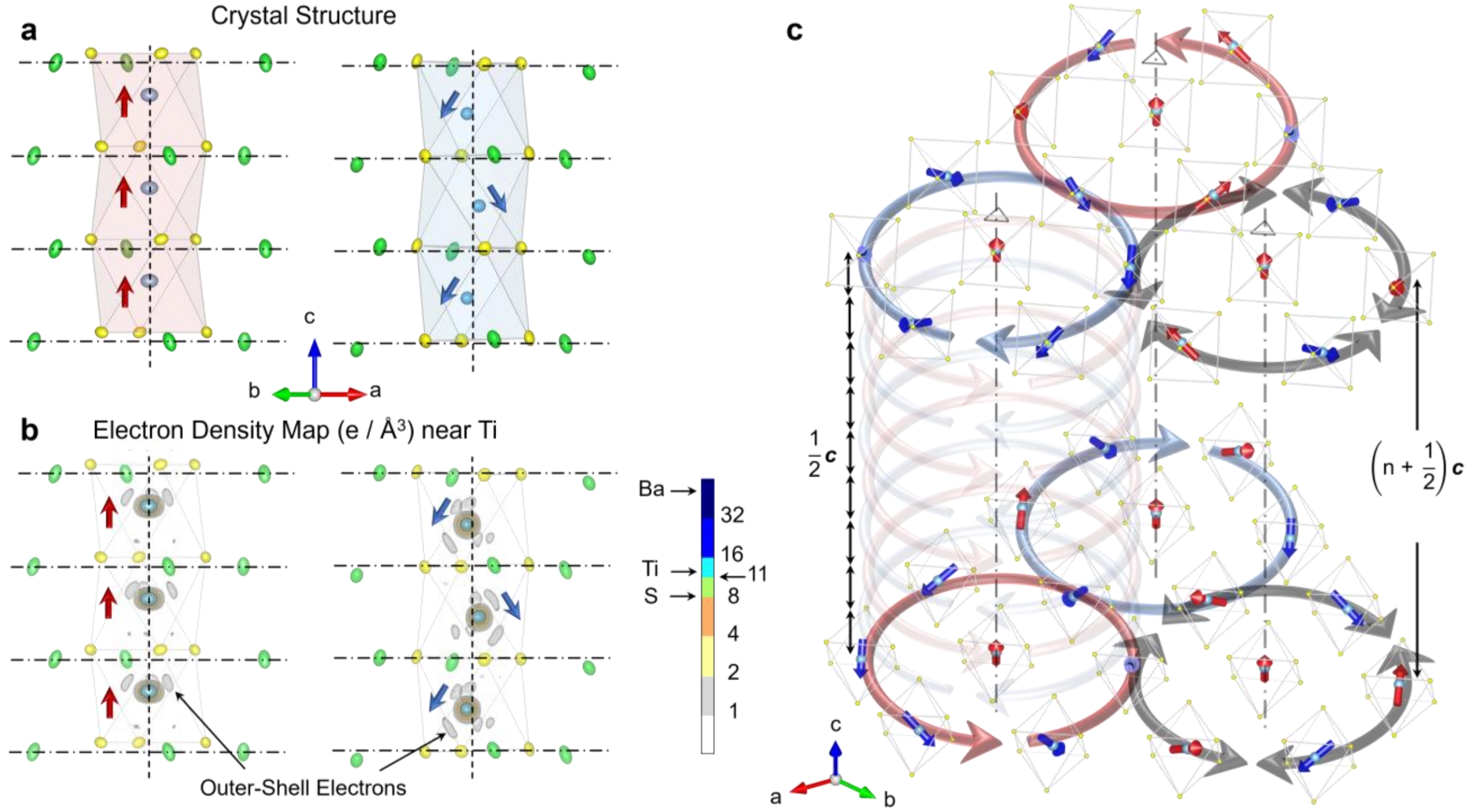

**Extended Data Figure 6. The dipolar $TiS_6$ chains of *P*3*c*1 $BaTiS_3$ with *c*-glide.** (**a**) and (**b**) show the same $TiS_6$ chains, one without and one with *a*-*b* plane dipolar displacements, (**a**) visualises the off-centred face-shared $TiS_6$ octahedra, and (**b**) focuses on the asymmetric electron density distributions near Ti. (**c**) Illustration of the *c*-glide symmetry of $BaTiS_3$. The *a*-*b* plane displacements glide every half unit cell along the *c*-axis, reversing the handedness between adjacent layers. This gives a minimum net handedness in the bulk $BaTiS_3$.

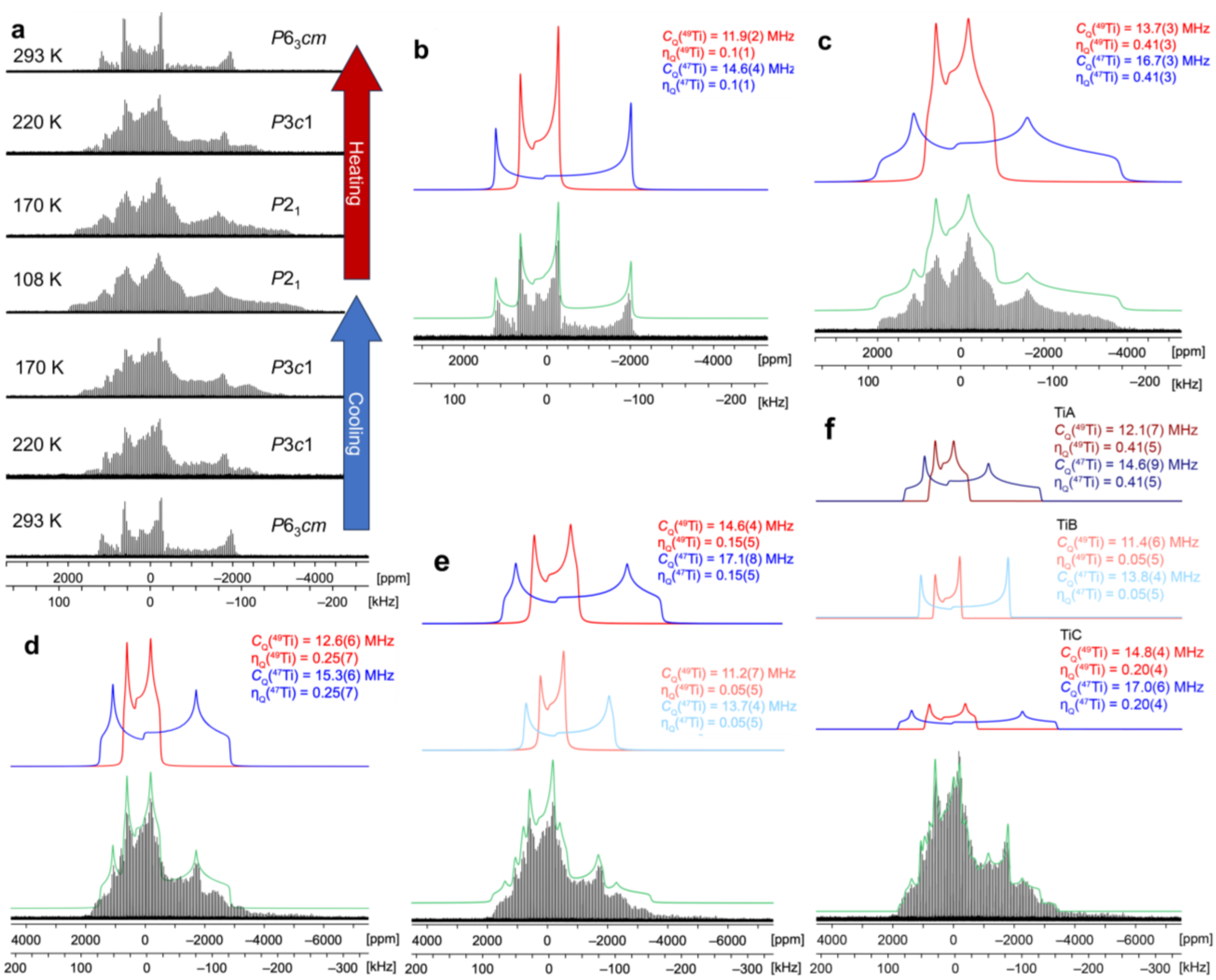

**Extended Data Figure 7. $^{47/49}$Ti SSNMR spectra of $BaTiS_3$**, (**a**) acquired at 18.8 T under static conditions at varying temperatures. Space groups of the three phases of $BaTiS_3$ observed upon cooling and heating are displayed to the right of each spectrum. Deconvolutions of experimental $^{47/49}$Ti SSNMR spectra of $BaTiS_3$ acquired at (**b**) 293 K (space group: $P6_3cm$), (**c**) 108 K (space group: $P2_1$), and (**d-f**) 170 K (space group: $P3c1$). Deconvolutions pictured in (**d**), (**e**), and (**f**) include spectra corresponding to one ($Ti^A$), two ($Ti^A$, $Ti^B$), and three ($Ti^A$, $Ti^B$, $Ti^C$) unique Ti environments, respectively. It can be seen in the spectra acquired at 293 K and 108 K that there exists only one unique Ti environment in the $P6_3cm$ and $P2_1$ phase, respectively. For the spectrum acquired at 170 K (cooling), a fit assuming three unique Ti environments in the $P3c1$ phase yields the best fit compared to those assuming one and two unique Ti environments, suggesting the presence of three distinguishable environments in this phase.

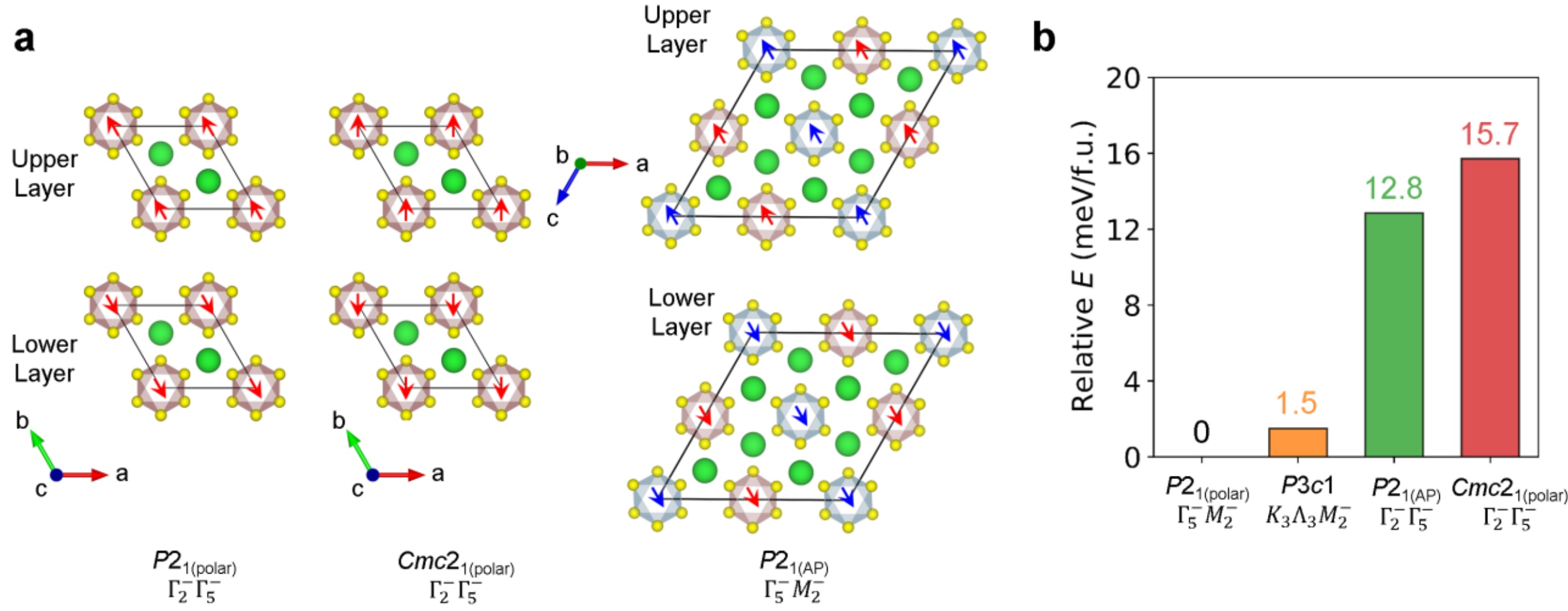

**Extended Data Figure 8. Energy comparison of $BaTiS_3$ phases.** (a) Atomic structures of different $BaTiS_3$ phases with their corresponding space groups and primary distortion mode irreducible representations (irreps). The structures include: a $1 \times 1 \times 1$ unit cell $P2_{1(\mathrm{polar})}$ with $\Gamma_2^-$ and $\Gamma_5^-$; a $1 \times 1 \times 1$ unit cell $Cmc2_{1(\mathrm{polar})}$ with $\Gamma_2^-$ and $\Gamma_5^-$; and a $2 \times 2 \times 1$ unit cell $P2_{1(\mathrm{antipolar,\ AP})}$ with $\Gamma_5^-$ and $\mathrm{M}_2^-$. Note that the $2\sqrt{3} \times 2\sqrt{3} \times 1$ unit cell $P3c1$ with $K_3$, $\Lambda_3$, and $M_2^-$ is not shown here but is included in **Figure 2c**. Ti atoms are displacing away from the centre of the $S_6$ octahedra. Out-of-plane displacement along the *c*-axis is indicated by colour (blue: downward, red: upward), while *a*-*b* plane displacements are shown with arrows. (b) Relative energies per formula unit (meV/f.u.) referenced to the ground-state $P2_{1(\mathrm{AP})}$ with $\Gamma_5^-$ and $\mathrm{M}_2^-$.

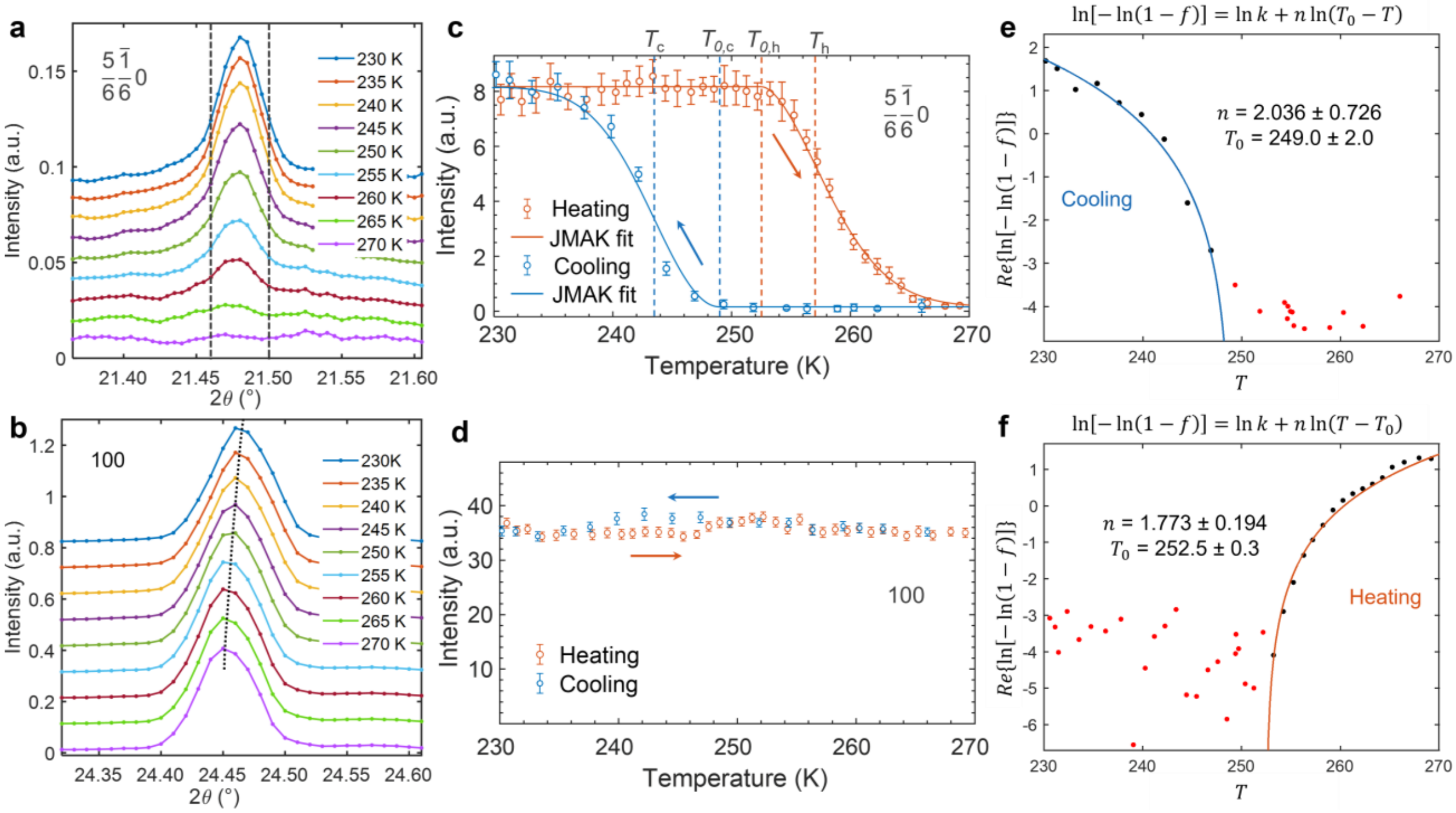

**Extended Data Figure 9. Temperature-dependent XRD of the nucleation and dissolution of the $BaTiS_3$ dipole vortices**. (**a-b**) Temperature-dependent XRD (heating) of (**a**) $\frac{5}{6}\frac{\bar{1}}{6}0$ and **(b)** 100 between 230 K and 270 K. The $\frac{5}{6}\frac{\bar{1}}{6}0$ Bragg intensity is extracted as the total intensity within the region of interest indicated by the dashed lines in (**a**). Lattice expansion across the CDW phase transition is highlighted by the dotted line in (**b**), with no clear sign of lattice size discontinuity. (**c-d**) Bragg intensities of (**c**) $\frac{5}{6}\frac{\bar{1}}{6}0$ and (**d**) 100 during cooling (blue) and heating (red). Experimental data are marked as circular dots with statistical error bars, while KJMA fitting is shown as solid lines. The hysteresis temperature values noted in (**c**) are nucleation temperatures $T_{0,c} = 249.0$ K, $T_{0,h} = 252.5$ K, inflexion temperatures $T_c = 243.2$ K, $T_h = 257.4$ K. The intensity of 100 in (**d**) remains relatively stable across the transition. (**e-f**) JKMA model fitting of the $\frac{5}{6}\frac{\bar{1}}{6}0$ intensity during consecutive (**e**) cooling and (**f**) heating, represented by blue and orange solid lines. Equilibrium XRD intensities before the transition happen are excluded from such fitting and are coloured as red data points.

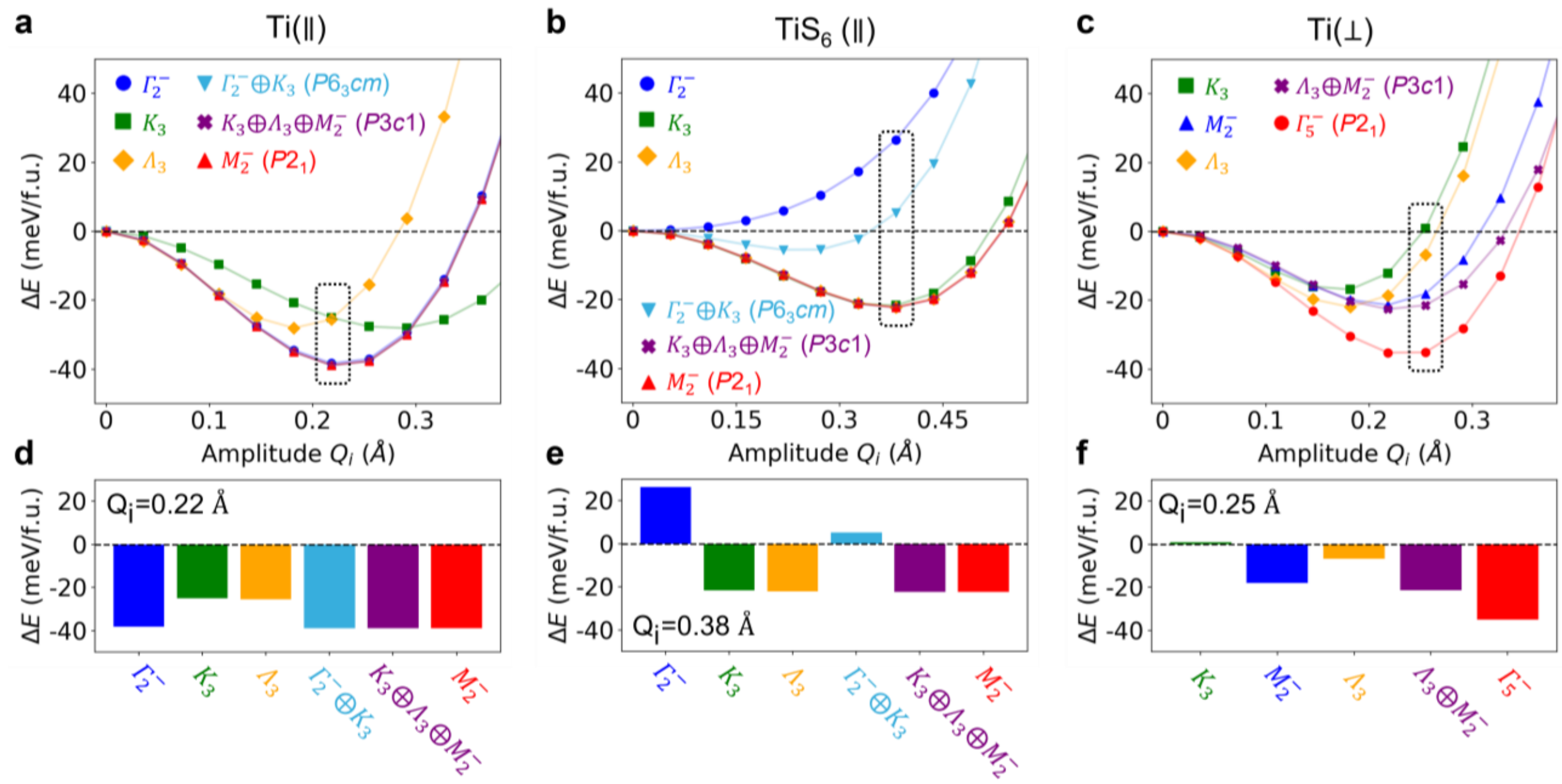

**Extended Data Figure 10. Evolution of the potential energy curves (PECs) with selected displacement modes involving single modes and their coupled forms in the experimentally observed phases** (*i.e.*, 130 K lamellar antiferroelectric phase $P2_1$, 220 K CDW-like antiferroelectric phase $P3c1$, and 300 K hexagonal ferrielectric phase $P6_3cm$). PECs for the modes related to (**a**) Ti off-centring along the *c*-axis (denoted as '(Ti ∥)'), (**b**) $TiS_6$ chain motion along the *c*-axis (denoted as '($TiS_6$ ∥)'), and (**c**) *a*-*b* plane Ti displacements (denoted as '(Ti ⊥)'). The black dotted box highlights the amplitude used for the comparison of energy shown in the bottom panels. (**d-f**) Bar plots depicting the energy values with each mode and their coupled forms in the observed phases at fixed amplitude ($Q_i$) corresponding to the minima in the PECs. (**d**) Energies of the (Ti ∥) modes at $Q_i$ = 0.22 Å. (**e**) Energies of the ($TiS_6$ ∥) modes at $Q_i$ = 0.38 Å. (**f**) Energies of the (Ti ⊥) modes at $Q_i$ = 0.25 Å.

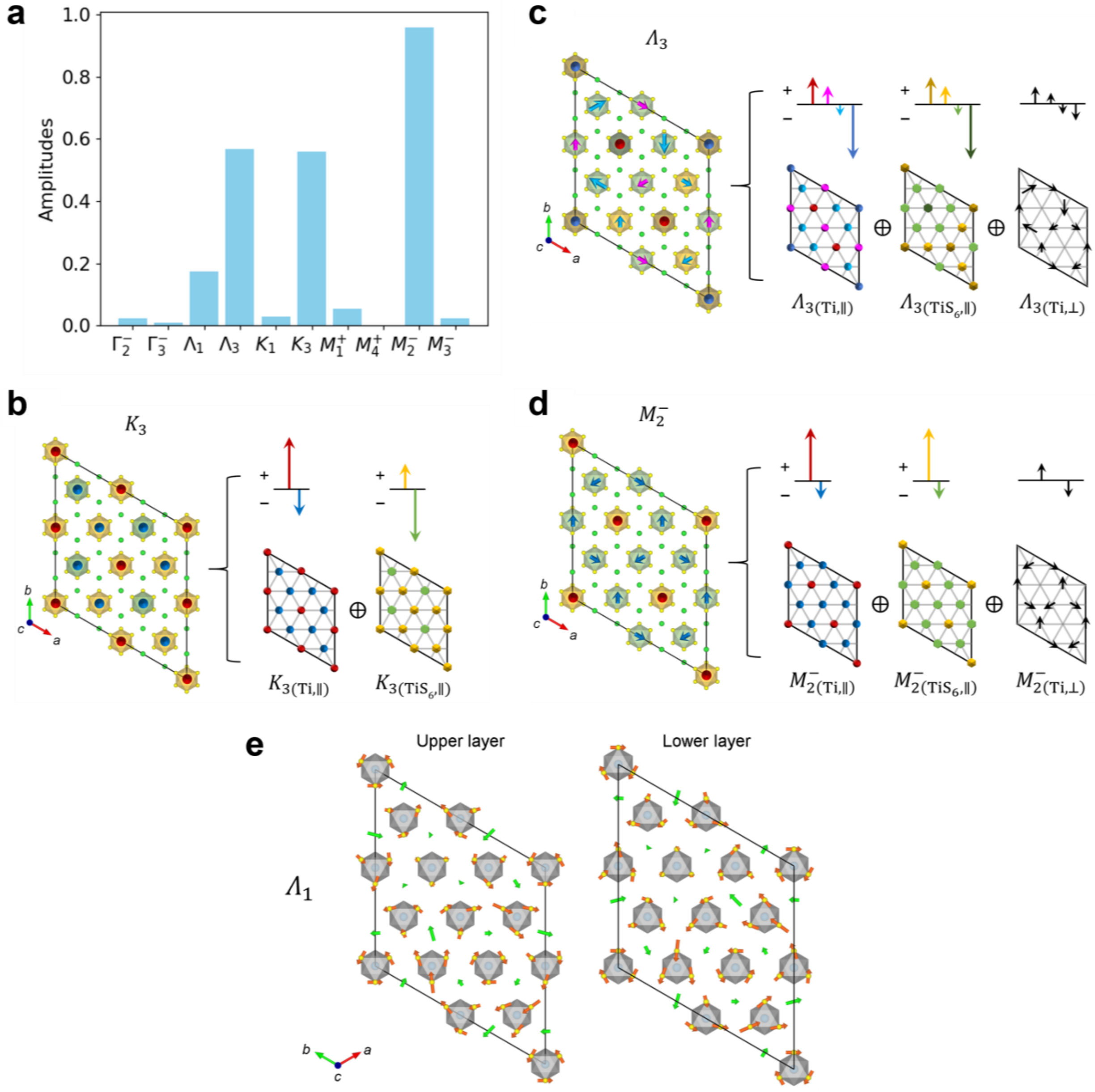

**Extended Data Figure 11. Schematic illustration of primary displacement modes in the *P*3*c*1 phase.** (**a**) Amplitudes of the displacement modes from centrosymmetric $P6_3/mmc$ $BaTiS_3$ to $P3c1$ $BaTiS_3$, identifying three primary and one secondary displacement modes, (**b**) $K_3$, (**c**) $\Lambda_3$, (**d**) $M_2^-$ and (**e**) $\Lambda_1$. Each of the primary displacement modes consists of up to three basis modes: Ti off-centring along the c-axis $(\mathrm{Ti}, \parallel)$, $TiS_6$ chain motions along the *c*-axis $(\mathrm{TiS_6}, \parallel)$, and *a*-*b* plane Ti displacements $(\mathrm{Ti}, \perp)$. $(\mathrm{Ti}, \parallel)$ motions are represented by blue (downward) and red (upward) colours at the centres of the $TiS_6$ polyhedron. $(\mathrm{TiS_6}, \parallel)$ motions are indicated by green (downward) and yellow (upward) colours at the $TiS_6$ polyhedron.

$(\mathrm{Ti}, \perp)$ motions are shown by black arrows. The upper panels show the relative amplitude at each site. Only the upper layers in the unit cell are shown; note that the lower layer has the opposite pattern of $(\mathrm{Ti}, \perp)$ but the same displacement along the c-axis: $(\mathrm{Ti}, \parallel)$ and $(\mathrm{TiS_6}, \parallel)$. (**e**) Schematic illustration of the $\Lambda_1$ mode. The S and Ba displacements along the *a*-*b* plane are shown by orange and green arrows, respectively.

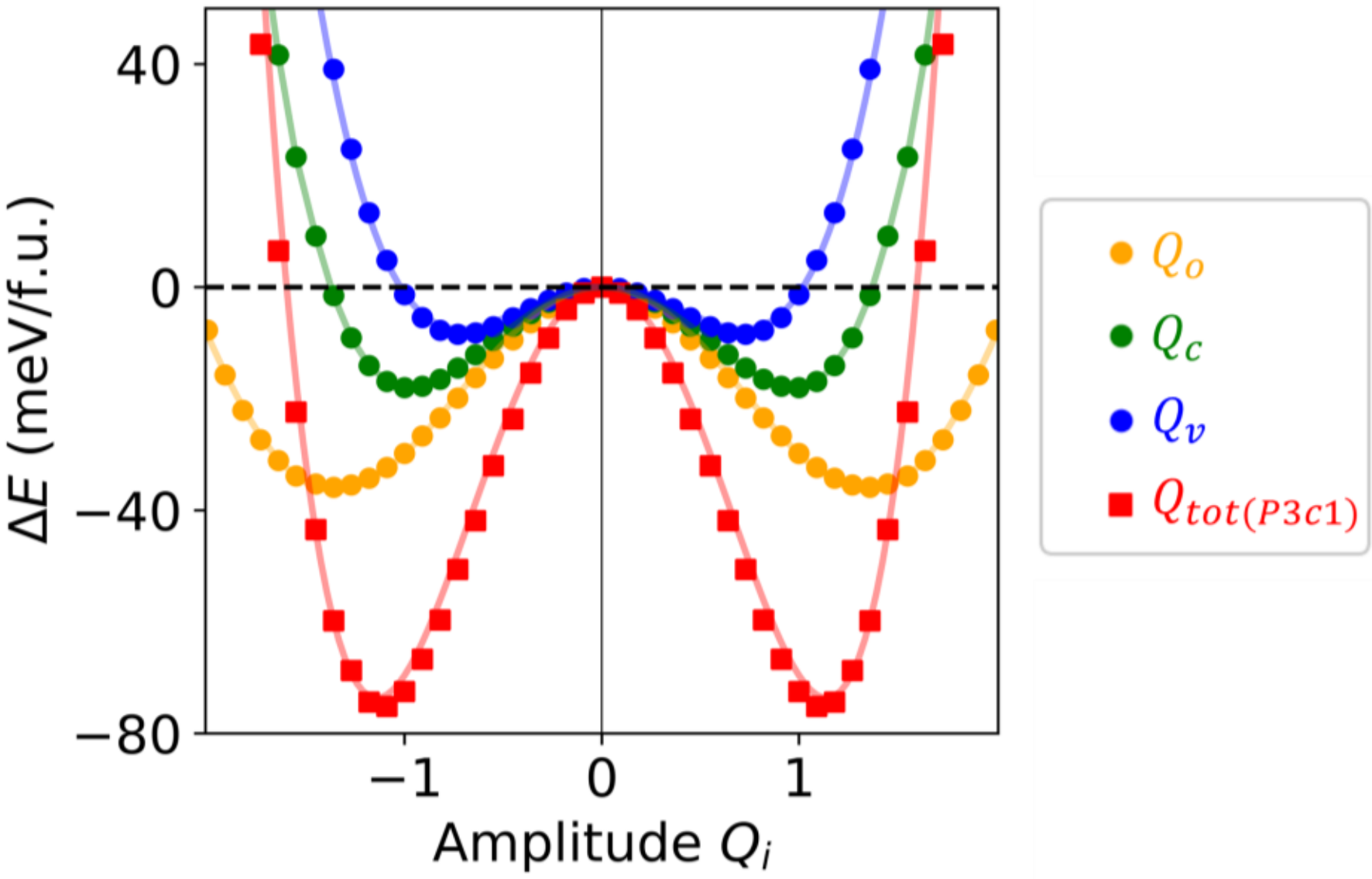

**Extended Data Figure 12. Fit of the DFT energies, represented by coloured dots, with the fitting curves for individual modes and total amplitude for *P*3*c*1, expressed in Eqn. 28,** including Ti off-centring along the *c*-axis (labelled as $Q_o$), $TiS_6$ chain motion along the *c*-axis (labelled as $Q_c$), and *a*-*b* plane Ti displacements (labelled as $Q_v$). The path connects the centrosymmetric $P6_3/mmc$ phase to the *P*3*c*1 phase. The amplitude $Q_i = 0$ corresponds to the $P6_3/mmc$ phase, and 1 corresponds to the *P*3*c*1 phase.

**Extended Data Table 1. Statistics of SC-XRD.** Data collection, unit cell determination, image scaling and integration, refinement file preparation, and refinement results in determining the crystal structure of 220K-$BaTiS_3$.

| | Rigaku, USC | | Bruker, ALS[31] | | High flux, Bruker, ALS* | |
|---|---|---|---|---|---|---|
| **Diffractometer** | XtaLAB Synergy-S | | Bruker D8 | | Bruker D8 | |
| **Source** | PhotonJet (Mo) X-ray Source | | Monochromatic Synchrotron | | Monochromatic Synchrotron | |
| **Crystal** | Platelet Crystal #1 | | Platelet Crystal #2 | | Platelet Crystal #1 | |
| **Size (μm)** | 80 × 20 × 15 | | 300 × 100 × 20 | | 50 × 20 × 15 | |
| **Wavelength** | 0.71073 Å | | 0.7288 Å | | 0.7288 Å | |
| **Temperature** | 220K | | 220K | | 220K | |
| **Space group** | $P3c1$ | | $P3c1$ | | $P3c1$ | |
| **Unit Cell Determination** | | | | | | |
| a, b, c (Å) | 23.3414(10), 23.3414(10), 5.8453(2) | | 23.285(5), 23.285(5), 5.836(2) | | 23.303(5), 23.303(5), 5.854(2) | |
| α, β, γ (°) | 90, 90, 120 | | 90, 90, 120 | | 90, 90, 120 | |
| Volume ($Å^3$) | 2758.0(3) | | 2740.0(15) | | 2753.1(15) | |
| Density ($g/cm^3$) | 4.067 | | 4.093 | | 4.074 | |
| **Integration and Scaling Statistics** | | | | | | |
| Resolution (Å) | Inf- 0.80 | Inf- 0.69 | Inf. - 0.62 | 0.72 - 0.62 | Inf - 0.60 | 0.70 - 0.60 |
| Completeness | 99.9 % | 87.0 % | 100.0 % | 100.0 % | 100 % | 100.0 % |
| Redundancy | 10.5 | 8.6 | 34.92 | 19.01 | 28.14 | 17.42 |
| Mean $I / \Delta I$ | 17.34 | 16.35 | 23.64 | 10.65 | 47.73 | 26.97 |
| $R_{int}$ | 0.056 | 0.055 | 0.0684 | 0.1557 | 0.0552 | 0.0545 |
| $R_{\sigma}$ | 0.031 | 0.027 | 0.0201 | 0.0490 | 0.0132 | 0.0198 |
| **Refinement File Preparation** | | | | | | |
| $h_{min}$, $h_{max}$ | -29, 29 | | -37, 37 | | -38, 38 | |
| $k_{min}$, $k_{max}$ | -29, 29 | | -37, 37 | | -38, 38 | |
| $l_{min}$, $l_{max}$ | -7, 7 | | -8, 9 | | -9, 9 | |
| $R_{\sigma}$, $R_{int}$ | 0.0421, 0.0517 | | 0.0315, 0.0691 | | 0.0234, 0.0472 | |
| $\theta_{min}$, $\theta_{full}$, $\theta_{max}$ (°) | 2.666, 25.242, 26.362 | | 1.035, 25.930, 35.40 | | 1.035, 25.930, 37.397 | |
| Friedel coverage | 0.961 | | 0.996 | | 0.996 | |
| **Refinement Results** | | | | | | |
| Weighting | $w = 1/[\sigma^2(F_o^2) + (0.0217P)^2 + 48.6113P]$, where $P = (F_o^2 + 2F_c^2)/3$ | | | | | |
| Cut-off Resolution | 0.80 Å | | 0.62 Å | | 0.59 Å | |
| No. reflections | 3705 | | 7983 | | 8480 | |
| No. $I > 2\sigma I$ | 3028 | | 6621 | | 8803 | |

| No. parameters | 183 | 185 | 185 |
| --- | --- | --- | --- |
| No. constraints | 1 | 1 | 1 |
| $R_1$, $wR_2$ [all data] | 0.0401, 0.0755 | 0.0420, 0.0692 | 0.0202, 0.0429 |
| $R_1$, $wR_2$ [$I > 4\sigma I$] | 0.0301, 0.0716 | 0.0305, 0.0657 | 0.0180, 0.0433 |
| *GoF* | 1.024 | 1.058 | 1.081 |
| Merohedral Twin | $\begin{pmatrix}1 & 0 & 0\\0 & 1 & 0\\0 & 0 & 1\end{pmatrix}$, $\begin{pmatrix}0 & -1 & 0\\-1 & 0 & 0\\0 & 0 & -1\end{pmatrix}$ | $\begin{pmatrix}1 & 0 & 0\\0 & 1 & 0\\0 & 0 & 1\end{pmatrix}$, $\begin{pmatrix}0 & -1 & 0\\-1 & 0 & 0\\0 & 0 & -1\end{pmatrix}$, $\begin{pmatrix}-1 & 0 & 0\\0 & -1 & 0\\0 & 0 & -1\end{pmatrix}$, $\begin{pmatrix}0 & 1 & 0\\1 & 0 & 0\\0 & 0 & 1\end{pmatrix}$ | $\begin{pmatrix}1 & 0 & 0\\0 & 1 & 0\\0 & 0 & 1\end{pmatrix}$, $\begin{pmatrix}0 & -1 & 0\\-1 & 0 & 0\\0 & 0 & -1\end{pmatrix}$, $\begin{pmatrix}-1 & 0 & 0\\0 & -1 & 0\\0 & 0 & -1\end{pmatrix}$, $\begin{pmatrix}0 & 1 & 0\\1 & 0 & 0\\0 & 0 & 1\end{pmatrix}$ |
| BASF | 0.49756 0.50244 | 0.39520 0.09971 0.10131 0.40378 | 0.22928 0.22604 0.27603 0.26865 |

*The following *P*3*c*1-$BaTiS_3$ crystal structure is derived from this run.

**Extended Data Table 2. Refinement results of 220K-$BaTiS_3$**: crystal structure.

| Atom | *x*/*a* | *y*/*b* | *z*/*c* | $U_{iso}$ (Å$^2$) | Occupancy | Symmetry Order |
|---|---|---|---|---|---|---|
| Ba01 | 0.33486(3) | 0.33313(3) | 0.45347(4) | 0.01281(5) | 1 | 1 |
| Ba02 | 0.33211(3) | 0.49896(2) | 0.89879(4) | 0.01467(8) | 1 | 1 |
| Ba03 | 0.66591(3) | 0.49955(2) | 0.43722(8) | 0.01365(10) | 1 | 1 |
| Ba04 | 0.16570(2) | 0.16753(3) | -0.07652(17) | 0.01573(16) | 1 | 1 |
| Ti01 | 0.33300(5) | 0.17224(3) | -0.23172(15) | 0.01335(9) | 1 | 1 |
| Ti02 | 0.49223(5) | 0.49201(5) | 0.1829(3) | 0.0173(2) | 1 | 1 |
| Ti03 | 0.333333 | 0.666667 | 1.5658(2) | 0.01638(17) | 1 | 3 |
| Ti04 | 0 | 0 | 0.5966(4) | 0.0189(4) | 1 | 3 |
| Ti05 | 0.15815(5) | 0.33351(7) | -0.32352(19) | 0.01406(16) | 1 | 1 |
| Ti06 | 2/3 | 1/3 | 0.6030(3) | 0.0151(2) | 1 | 3 |
| S00A | 0.08175(8) | 0.24953(9) | -0.0945(3) | 0.0112(3) | 1 | 1 |
| S00B | 0.16529(10) | 0.41637(8) | -0.0920(3) | 0.0107(3) | 1 | 1 |
| S00C | 0.08395(7) | 0.08209(7) | 0.3728(4) | 0.0102(3) | 1 | 1 |
| S00D | 0.08494(7) | 0.33283(11) | -0.6072(3) | 0.0113(2) | 1 | 1 |
| S00E | 0.58337(6) | 0.33308(11) | 0.3829(3) | 0.0105(2) | 1 | 1 |
| S00F | 0.41871(10) | 0.50140(11) | 0.4573(4) | 0.0138(4) | 1 | 1 |
| S00G | 0.50038(10) | 0.41810(9) | 0.4581(3) | 0.0113(3) | 1 | 1 |
| S00H | 0.25125(9) | 0.16634(10) | -0.5124(3) | 0.0105(3) | 1 | 1 |
| S00I | 0.58483(8) | 0.58471(7) | 0.4451(4) | 0.0140(3) | 1 | 1 |
| S00J | 0.33222(11) | 0.25068(7) | -0.0029(2) | 0.01090(16) | 1 | 1 |
| S00K | 0.41672(9) | 0.24858(9) | -0.5094(3) | 0.0102(3) | 1 | 1 |
| S00L | 0.33417(11) | 0.58397(6) | 0.3428(2) | 0.0116(2) | 1 | 1 |

**Extended Data Table 3. Refinement results of 220K-$BaTiS_3$**: anisotropic atomic displacement parameters.

| Atom | $U_{11}$ (Å$^2$) | $U_{22}$ (Å$^2$) | $U_{33}$ (Å$^2$) | $U_{23}$ (Å$^2$) | $U_{13}$ (Å$^2$) | $U_{12}$ (Å$^2$) |
|---|---|---|---|---|---|---|
| Ba01 | 0.01135(9) | 0.01155(10) | 0.01543(13) | -0.00249(14) | 0.00102(18) | 0.00565(14) |
| Ba02 | 0.01155(10) | 0.01161(14) | 0.01989(18) | -0.00373(13) | 0.0016(2) | 0.00506(13) |
| Ba03 | 0.01116(9) | 0.01255(14) | 0.0174(3) | 0.00578(14) | 0.0013(2) | 0.00607(13) |
| Ba04 | 0.01225(16) | 0.01087(14) | 0.0241(5) | 0.00057(15) | 0.00560(16) | 0.00582(12) |
| Ti01 | 0.0177(2) | 0.0163(4) | 0.0078(2) | -0.0010(3) | 0.0005(4) | 0.0098(3) |
| Ti02 | 0.0168(5) | 0.0149(5) | 0.0159(6) | 0.0005(3) | 0.0010(4) | 0.0046(3) |
| Ti03 | 0.0206(3) | 0.0206(3) | 0.0079(4) | 0 | 0 | 0.01030(13) |
| Ti04 | 0.0215(5) | 0.0215(5) | 0.0136(9) | 0 | 0 | 0.0108(2) |
| Ti05 | 0.0144(5) | 0.0204(3) | 0.0106(4) | -0.0002(5) | 0.0000(3) | 0.0111(4) |
| Ti06 | 0.0204(3) | 0.0204(3) | 0.0045(7) | 0 | 0 | 0.01020(15) |
| S00A | 0.0087(4) | 0.0104(4) | 0.0118(7) | -0.0028(5) | -0.0013(4) | 0.0028(4) |
| S00B | 0.0121(5) | 0.0081(4) | 0.0127(8) | -0.0017(4) | 0.0004(5) | 0.0056(4) |
| S00C | 0.0107(5) | 0.0085(5) | 0.0092(6) | 0.0010(4) | -0.0009(5) | 0.0030(3) |
| S00D | 0.0080(4) | 0.0120(3) | 0.0150(6) | 0.0011(6) | -0.0019(4) | 0.0057(4) |
| S00E | 0.0070(5) | 0.0137(3) | 0.0109(5) | -0.0030(7) | -0.0015(5) | 0.0054(5) |
| S00F | 0.0096(5) | 0.0135(6) | 0.0185(9) | -0.0011(6) | 0.0011(4) | 0.0059(4) |
| S00G | 0.0126(5) | 0.0095(5) | 0.0137(6) | 0.0002(5) | 0.0010(5) | 0.0069(4) |
| S00H | 0.0077(5) | 0.0149(6) | 0.0100(8) | -0.0023(5) | -0.0032(4) | 0.0064(4) |
| S00I | 0.0114(5) | 0.0104(5) | 0.0179(9) | 0.0018(4) | 0.0016(5) | 0.0038(4) |
| S00J | 0.0145(3) | 0.0089(4) | 0.0106(4) | -0.0003(3) | 0.0001(6) | 0.0068(4) |
| S00K | 0.0097(5) | 0.0083(5) | 0.0104(7) | -0.0005(5) | -0.0020(4) | 0.0029(4) |
| S00L | 0.0143(3) | 0.0107(5) | 0.0129(6) | -0.0005(4) | -0.0008(8) | 0.0086(5) |

**Extended Data Table 4. The Ti off-centring displacement of the CDW-$BaTiS_3$ at 220 K and the resulting asymmetric quadrupolar parameters in DFT**: Ti off-centring concerning the $S_6$ octahedron are expressed as $\Delta x/a$, $\Delta y/b$, and $\Delta z/c$. The asymmetric Ti–S coordination at each Ti site results in nonzero Ti quadrupolar parameters, $C_Q$ (of 47Ti and 49Ti) and $\eta_Q$, as determined by DFT.

| Atom | Displacement Vector | | | Magnitude (Å) | | Titanium Quadrupolar Parameters | | |
|---|---|---|---|---|---|---|---|---|
| | $\Delta x/a$ | $\Delta y/b$ | $\Delta z/c$ | $\perp$ ***c*** | $\parallel$ ***c*** | $C_Q(^{49}Ti)$ (MHz) | $C_Q(^{47}Ti)$ (MHz) | $\eta_Q$ |
| Ti01 | -0.00038 | 0.00555 | 0.02651 | 0.134 | 0.155 | 11.50 | 14.08 | 0.43 |
| Ti02 | -0.00772 | -0.00804 | -0.02060 | 0.184 | 0.121 | 12.13 | 14.85 | 0.46 |
| Ti03 | 0.00000 | 0.00000 | -0.02700 | 0.000 | 0.158 | 10.22 | 12.51 | 0 |
| Ti04 | 0.00000 | 0.00000 | -0.02620 | 0.000 | 0.153 | 10.34 | 12.65 | 0 |
| Ti05 | -0.00835 | 0.00032 | 0.02438 | 0.199 | 0.143 | 12.68 | 15.52 | 0.17 |
| Ti06 | 0.00000 | 0.00000 | -0.02990 | 0.000 | 0.175 | 9.70 | 11.88 | 0 |

**Extended Data Table 5. Experimentally determined quadrupolar parameters for each temperature.**

| Titanium Site | $\delta_{iso}(^{49}Ti)$ (ppm)[a] | $\delta_{iso}(^{47}Ti)$ (ppm)[a] | $C_Q(^{49}Ti)$ (MHz)[b] | $C_Q(^{47}Ti)$ (MHz)[b] | $\eta_Q(^{49}Ti)$[c] | $\eta_Q(^{47}Ti)$[c] |
|---|---|---|---|---|---|---|
| 293 K (cooling) – 1 site – *P*$6_3$*cm* | | | | | | |
| Ti1 | 310 | 326 | 11.9(2) | 14.6(4) | 0.1(1) | 0.015(15) |
| 170 K (cooling) – 1 site (220 K cooling looks similar, but poorer S/N) – *P*3*c*1 | | | | | | |
| TiA | 300 | 266 | 12.6(6) | 15.3(6) | 0.25(7) | 0.25(3) |
| 170 K (cooling) – 2 sites – *P*3*c*1 | | | | | | |
| TiA | 340 | 266 | 14.6(4) | 17.1(8) | 0.15(5) | 0.25(25) |
| TiB | 310 | 325 | 11.2(7) | 13.7(4) | 0.05(5) | 0.05(5) |
| 170 K (cooling) – 3 sites – *P*3*c*1 – BEST FIT | | | | | | |
| TiA | 350 | 350 | 12.1(7) | 14.6(9) | 0.41(5) | 0.41(4) |
| TiB | 310 | 306 | 11.4(6) | 13.8(4) | 0.05(5) | 0.03(3) |
| TiC | 340 | 256 | 14.8(4) | 17.0(6) | 0.20(4) | 0.20(4) |
| 108 K (cooling) – 1 site – *P*$2_1$ | | | | | | |
| Ti1 | 276 | 276 | 13.7(3) | 16.7(3) | 0.41(3) | 0.41(3) |
| 220 K (heating) – 2 sites (170 K heating looks similar, but poorer S/N) – 2 sites – *P*3*c*1<br>**N.B.:** We could not get a reasonable 3-site fit here due to poorer signal-to-noise. | | | | | | |
| Ti1 | 300 | 376 | 12.5(4) | 15.5(2) | 0.20(6) | 0.20(3) |
| Ti2 | 326 | 326 | 11.3(7) | 13.9(5) | 0.05(5) | 0.05(5) |
| 293 K (heating) – *P*$6_3$*cm* | | | | | | |
| Ti1 | 320 | 326 | 11.9(2) | 14.5(2) | 0.05(5) | 0.02(2) |

[a] Uncertainties in chemical shifts are estimated as ±40, ±100, and ±20 ppm for spectra corresponding to the *P*$6_3$*cm*, *P*3*c*1, and *P*$2_1$ phases, respectively.

[b] The quadrupolar coupling constants, $C_Q(^{47}Ti)$ and $C_Q(^{49}Ti)$, represent best fits. Ideally, the ratio of $C_Q(^{47}Ti)/C_Q(^{49}Ti) = Q(^{47}Ti)/Q(^{49}Ti) = 1.22$, where Q are the nuclear quadrupole moments of $^{47}Ti$ and $^{49}Ti$ – these conditions are limited by the uncertainties in the fits, but the fits honour these conditions in the limits of uncertainty.

[c] The asymmetry parameters, $\eta_Q(^{49}Ti)$ and $\eta_Q(^{47}Ti)$, represent best fits. Ideally, $\eta_Q(^{49}Ti)$ and $\eta_Q(^{47}Ti)$ should be equal – these conditions are limited by the uncertainties in the fits, but the fits honour these conditions in the limits of uncertainty.

**Extended Data Table 6**. Landau coefficients of the free energy expansion, truncated at sixth order.

| $\alpha_1$ (meV/f.u./Å$^2$) | $\alpha_2$ (meV/f.u./Å$^4$) | $\alpha_3$ (meV/f.u./Å$^6$) | $\beta_1$ (meV/f.u./Å$^2$) | $\beta_2$ (meV/f.u./Å$^4$) | $\beta_3$ (meV/f.u./Å$^6$) | $\gamma_1$ (meV/f.u./Å$^2$) | $\gamma_2$ (meV/f.u./Å$^4$) |
|---|---|---|---|---|---|---|---|
| -44.21 | 15.57 | -1.26 | -36.71 | 26.64 | -0.67 | -30.02 | 30.92 |
| $\gamma_3$ (meV/f.u./Å$^6$) | $\delta_1$ (meV/f.u./Å$^4$) | $\delta_2$ (meV/f.u./Å$^4$) | $\delta_3$ (meV/f.u./Å$^6$) | $\delta_4$ (meV/f.u./Å$^6$) | $\delta_5$ (meV/f.u./Å$^6$) | $\delta_6$ (meV/f.u./Å$^6$) | $\epsilon_1$ (meV/f.u./Å$^4$) |
| -2.01 | 3.84 | 0.32 | -22.47 | 22.43 | 144.53 | -144.01 | 29.79 |
| $\epsilon_2$ (meV/f.u./Å$^4$) | $\epsilon_3$ (meV/f.u./Å$^6$) | $\epsilon_4$ (meV/f.u./Å$^6$) | $\epsilon_5$ (meV/f.u./Å$^6$) | $\epsilon_6$ (meV/f.u./Å$^6$) | $\eta_1$ (meV/f.u./Å$^4$) | $\eta_2$ (meV/f.u./Å$^4$) | $\eta_3$ (meV/f.u./Å$^6$) |
| -34.42 | -24.66 | 22.81 | 144.90 | -142.30 | -5.41 | -19.23 | -130.75 |
| $\eta_4$ (meV/f.u./Å$^6$) | $\eta_5$ (meV/f.u./Å$^6$) | $\eta_6$ (meV/f.u./Å$^6$) | $\kappa_1$ (meV/f.u./Å$^4$) | $\kappa_2$ (meV/f.u./Å$^4$) | $\kappa_3$ (meV/f.u./Å$^4$) | $\kappa_4$ (meV/f.u./Å$^6$) | $\kappa_5$ (meV/f.u./Å$^6$) |
| 130.54 | -50.62 | 52.69 | -4.00 | -12.19 | 8.11 | 2.16 | 5.10 |
| $\kappa_6$ (meV/f.u./Å$^6$) | $\kappa_7$ (meV/f.u./Å$^6$) | $\kappa_8$ (meV/f.u./Å$^6$) | $\kappa_9$ (meV/f.u./Å$^6$) | $\kappa_{10}$ (meV/f.u./Å$^6$) | $\kappa_{11}$ (meV/f.u./Å$^6$) | $\kappa_{12}$ (meV/f.u./Å$^6$) | $\kappa_{13}$ (meV/f.u./Å$^6$) |
| -173.01 | 167.27 | -25.21 | 156.54 | -67.40 | -88.98 | -43.60 | 69.34 |